\documentclass[manuscript]{aastex}

\newcommand{\be}{\begin{equation}}
\newcommand{\ee}{\end{equation}}
\newcommand{\bea}{\begin{eqnarray}}
\newcommand{\eea}{\end{eqnarray}}

\slugcomment{}

\shorttitle{Radiative properties of magnetized neutron star surfaces}
\shortauthors{Suleimanov et al.}

\begin{document}


\title{Radiative properties of highly magnetized isolated \\
  neutron star surfaces and approximate treatment \\ 
of absorption features in their spectra    }


\author{V. Suleimanov\altaffilmark{1,2}, V.V. Hambaryan\altaffilmark{3},
  A.Y. Potekhin\altaffilmark{4,5,6}, M. van Adelsberg\altaffilmark{7},
  R.~Neuh\"auser\altaffilmark{3}, and  K. Werner\altaffilmark{1}}
\affil{}
\email{}


\altaffiltext{1}{Institute for Astronomy and Astrophysics, Kepler Center for Astro and Particle Physics,
   Eberhard Karls University, Sand 1, 72076 T\"ubingen, Germany; suleimanov@astro.uni-tuebingen.de}
\altaffiltext{2}{Dept. of Astronomy, Kazan State University, Kremlevskaya 18, 420008 Kazan, Russia}
\altaffiltext{3}{Astrophysikalisches Institut und Universit\"ats-Sternwarte Jena, Schillerg\"asschen 2-3, 07745 Jena,
 Germany}
\altaffiltext{4}{Ioffe Physical-Technical Institute,  Politekhnicheskaya str., 26, St.
Petersburg 194021, Russia}
\altaffiltext{5}{CRAL (UMR CNRS No.\,5574), Ecole Normale Sup\'erieure de Lyon,
69364 Lyon Cedex 07, France}
\altaffiltext{6}{Isaac Newton Institute of Chile, St.~Petersburg Branch, Russia}
\altaffiltext{7}{Kavli Institute for Theoretical Physics, Kohn Hall, University of California, Santa Barbara, CA 93106, USA
}


\begin{abstract}
In the X-ray spectra of most X-ray dim isolated neutron stars (XDINSs)  absorption features
with equivalent widths  (EWs) of 50 -- 200 eV are observed. These features are
usually connected with the proton
cyclotron line, but their nature is not well known yet.
We theoretically investigate different models to explain these absorption features 
and compare their properties with the observations for a better understanding of the radiation properties of magnetizedneutron star surfaces.  Based on these models, we create a fast and flexible
  code to fit observed spectra of isolated neutron stars. 
We consider various theoretical models for the magnetized neutron star surface: naked condensed iron
surfaces and partially ionized hydrogen model atmospheres,
including semi-infinite and thin atmospheres above a
condensed surface.  Spectra of condensed iron surfaces are
  represented by a simple analytical approximation.
The condensed surface radiation properties are considered as the
inner atmosphere boundary condition for the thin atmosphere.
The properties of the absorption features (especially equivalent widths) and the angular distributions of the
emergent radiation are described for all models. 
A code for computing light curves and integral emergent
spectra of magnetized neutron stars is developed. We assume a dipole surface magnetic field distribution
with a possible toroidal component and corresponding temperature distribution. A model with two uniform hot spots at the
magnetic poles can also be employed.
Light curves and spectra of highly magnetized neutron stars with
parameters typical for XDINSs
are computed using different surface temperature distributions and
 various local surface models.
 Spectra of magnetized model atmospheres are  approximated by diluted
  blackbody  spectra with one or two Gaussian lines having parameters, which allow us to describe
  the model absorption features.  
The EWs of the absorption
features in the integral spectra cannot exceed significantly 100 eV 
for a magnetic atmosphere or a
naked condensed surface as   local surface models. A thin atmosphere above
 a condensed surface can have an absorption feature whose EW
exceeds 200 eV in the integrated spectrum. If the toroidal component of the magnetic field on the
neutron star  atmosphere is  3 -- 7 times larger than the poloidal
component, the absorption feature in the integral spectrum is too
wide and shallow to be detectable.  
To explain the prominent absorption features in the soft X-ray spectra 
of XDINSs a thin atmosphere above the condensed
surface can be invoked, whereas a strong toroidal magnetic field component
 on the XDINS surfaces can be excluded.
\end{abstract}



\section{Introduction}
\label{s:intr}

X-ray dim isolated neutron stars (XDINSs) represent a new class of X-ray sources,
established by ROSAT observations \citep{Walteretal:96}. At present,
seven such objects have been found
(``The Magnificent Seven'',
see review by \citealt{Haberl:07}). All of these
objects have ages $\approx 10^6$ yr and thermal-like spectra with
effective temperatures $T_{\rm eff} \sim 10^6$ K. The radiation of XDINSs is pulsed with amplitudes
from 1\% (RX\,J1856.5$-$3754, \citealt{Tie.Meregh:07}) to 18\%
(RBS 1223, \citealt{Swopeetal:05}).

Absorption features in the 0.2\,--\,0.8 keV range
 are detected in the spectra of almost all XDINSs \citep{Haberl:07}.
In three objects the presence of two (RBS 1223, \citealt{Swopeetal:07}, and RX\,J0806.4-4123,
\citealt{Haberl:07}) or three (RX\,J1605.3+3249, \citealt{Haberl:07}) absorption lines is claimed.
Equivalent widths of these features are about 30\,--\,100 eV \citep{Haberletal:04,Haberl:07} with
the most prominent absorption in RBS 1223 ($\approx 200$ eV, \citealt{Swopeetal:07}).
If these features are interpreted as proton cyclotron lines, they correspond to
magnetic fields $B\sim$   0.4 - 1.6 $\times 10^{14}$ G on the XDINSs surfaces, but
if they are interpreted as electron cyclotron lines, then
$B\sim$  2 - 8 $\times 10^{10}$ G (see \citealt{vKK:07}).

The optical
counterparts of some XDINSs
have been found  (see reviews by \citealt{Mignanietal:07, Mignani:09} and
  references therein, and  \citealt{eis:10}).  It is very intriguing that their optical/ultraviolet fluxes are a few
times larger than the  blackbody extrapolation of the X-ray spectra
\citep{Burwitzetal:01,Burwitzetal:03,Kaplanetal:03,Motchetal:03}. Two explanations for this excess have
been suggested.
First, it can be caused by non-uniform temperature distributions across XDINS surfaces.
Second, the emergent spectral energy distribution of the surfaces can
differ from the blackbody. For example, using non-magnetic model atmosphere spectra gives an
even larger optical excess compared to the observed one \citep{pavlovetal:96}. \citet{Hoetal:07}
explained the optical excess in RX\,J1856.5$-$3754 by a single magnetic model atmosphere
of finite thickness above a condensed surface. Both of these
mechanisms can be important for XDINSs.

The XDINSs are nearby objects, and distances to some of them have been
estimated by parallax measurements \citep{Kaplanetal:02a,vKK:07}.
This helps to evaluate their
radii \citep{Trumperetal:04}, yielding useful information on the equation of state
(EOS) for the neutron-star core, which is important
for the studies of particle interactions at supranuclear densities and pressures
\citep{LP07,NSB1}
as well as for
computations of templates of gravitational wave signals which arise
during neutron stars merging (e.g. \citealt{Baiottietal:08}).

The theoretical emergent spectra of the XDINS surfaces are also necessary
for the evaluation of neutron star radii.
The XDINS surface layers can either be condensed or have plasma
atmospheres  (\citealt{Rud:71,Romani:87}, see also \citealt{Lai.Salpeter:97,
  Lai:01, ML:07}),
depending on surface temperature
and magnetic field strength. The radiation spectra of magnetized condensed surfaces
were calculated by many authors in the framework of different assumptions \citep{brin:80,tur:04,
PerAzetal:05,vAetal:05}.
Investigations of the neutron-star atmospheres are even more extended, from the
first relatively simple models \citep{Romani:87,Shibanovetal:92} to more sophisticated
models, including  complete \citep{Lai.Ho:02, Ho.Lai:03} or
partial \citep{Lai.Ho:03,vAL:06} photon mode conversion due to the vacuum polarization effect
\citep{PavlovGnedin},
and partially ionized hydrogen models
\citep{Potekhinetal:04,Hoetal:08}. Mid-$Z$ element partially ionized atmospheres for 
magnetized neutron stars (however without detailed account of atomic motion effects) have
also been modeled and used for the absorption features interpretation
(see \citealt{MH:07} and references therein). 

Absorption features in the soft X-ray spectra of XDINSs may be the
Rosetta stone for understanding of the physical nature of their surfaces.
For this purpose it is necessary to fit the observed spectra 
by theoretical spectra of model atmospheres and condensed surfaces.
A correct model
should allow one to describe the observed absorption features, as well as
the overall spectral energy distribution, including the optical/UV
fluxes, and the pulse profiles. For such a comparison an integration of local
spectra over the neutron star surface should be performed with relativistic effects taken into account.
Therefore, the distribution of the magnetic field and the effective temperature over the neutron
star surface has to be found also.  Computations of integral spectra from local 
atmosphere emergent spectra were carried out by many authors
(e.g., \citealt{Zavlin:95,Zan.Tur:06,Hoetal:08}). Some authors have also used
 model atmospheres  to fit the observed isolated
 XDINS X-ray spectra  (\citealt{Zan.Tur:06,Ho:07}, see also review
 by \citealt{Zavlin:09} and references therein).

Usually a simple dipole magnetic field distribution over the neutron star surface with corresponding
relativistic correction is assumed \citep{Pav.Zav:00,Hoetal:08}.  If this surface dipole field
corresponds to a global dipole field of the star,
one can find a surface effective temperature distribution depending on
the magnetic field strength and the inner neutron-star temperature \citep{Gren.Hart:83,
Pot.Yak:01,Potekhinetal:03}. The surface temperature distribution is defined by the local inclination angle
of the magnetic field to the surface normal, which determines the value of the local radial thermal
conductivity in the neutron star crust.  Bright pole spots are too large  to explain the observed
XDINSs' pulsed fractions if the dipole field model and the  blackbody
approximation are used for the emergent spectra \citep{Geppertal:06}. Quadrupole
or more complicated magnetic field configurations in neutron star crusts
have been suggested (e.g.,
\citealt{Ruderman91,Zan.Tur:06}), which may provide
an explanation for the high pulsed fraction,
although an equally likely explanation, possibly, may be provided
by models of polar caps heating.  \citet{Geppertal:04} assumed that the magnetic field is concentrated in the crust only.
\cite{Geppertal:06} and
\cite{PerAzetal:06a}
propose that the magnetic field in the crust additionally has a strong toroidal component.
In both cases the
polar hot spots are small and can explain the observed XDINS pulsations. In particular, \cite{PerAzetal:06b}
used the model with the toroidal field to interpret all the observed features of RX\,J0720$-$3125 on the base
of a condensed surface emission.

 At present time, most of the observed XDINSs spectra were fitted by a
simple blackbody model with one or more Gaussian lines. The detailed modeling of
observed spectra by using accurate model atmosphere or condensed surface spectra
is computationally very expensive, particularly if the temperature and magnetic field
distributions over the stellar surface are considered.  For example, the spectrum of
RX\,J1856.5$-$3754 was described by a single model atmosphere \citep{Hoetal:07},
and the pulsed  fraction of the same object was modeled by using only four
latitude points \citep{Ho:07}. This is one possible way for fitting  spectra of
XDINSs: Create an extensive set of theoretical XDINSs spectra with (inevitably)
limited variation of stellar parameters as well as temperature and magnetic
field distributions over the surface (see, e.g. \citealt{Hoetal:08}). We suggest
an alternative way. The local spectra of the neutron star  together with
temperature and magnetic field distributions can be fitted by simple analytical
functions. Then  an integral spectrum of the neutron star with a sufficiently
detailed latitude set can be computed rather quickly. This approach is more
flexible and faster, but the results will be more approximate.  However, in this
way one can first constrain the XDINS parameters, and then the more accurate,
expensive calculations can be performed to refine the analysis.  Here we develop this approximate
approach and its application for fitting XDINSs spectra will be presented in
a different paper.  Some preliminary conclusions about the possible nature of the
XDINSs emitting surface will be made on the base of test calculations.

 In this work we analyze the properties of emergent spectra of magnetized model atmospheres, condensed surfaces, and thin
magnetized model atmospheres above condensed surfaces, paying
special attention to the equivalent widths of the absorption features and the angular distribution of the emergent radiation.
 Spectra of naked condensed surfaces are  represented using a simple
  analytical approximation.
We also present test calculations of the integral spectra and
light curves of the magnetized neutron star models, using a simple
representation of the considered local model spectra  by analytical functions.
These calculations allow us
to formulate the  qualitative
characteristics of the local spectra that
are  necessary  to fit the observed pulsed fractions of XDINS light curves 
and the absorption features of XDINS spectra.  
These simple computations allow us also to choose the model of the radiative
  neutron star surface, which will be used later for more extensive and
  accurate computations of the integral spectra and the light curves of XDINSs.

\section {Local models of a neutron star surface}
\label{s:lmodel}

A magnetized neutron star surface can be condensed or can have  a plasma  atmosphere.
It depends mainly on chemical composition, temperature, and magnetic field strength, but
necessary conditions are not established exactly
\citep{Lai.Salpeter:97,PCS:99,Lai:01,PCh:04,ML:07}.
In this section we shortly describe
theoretical radiation properties of the condensed surfaces and the model atmospheres of
magnetized XDINSs. The main attention is concentrated on the equivalent widths of the absorption features
and the angular distributions of the emergent spectra, which are also very important for the integral
spectra computations. The model atmospheres are discussed in more detail. 

\subsection{Radiation from the condensed surface of magnetized neutron stars}

The description of magnetized condensed surface radiation properties is based on the results of
\citet{vAetal:05} (see their Figs. 2--6;
the results of \citealt{PerAzetal:05} are in good agreement).
They were obtained using a free-ion approximation, which can be not exactly
accurate. An earlier work by \citet{tur:04} was based on the approximation of fixed (non-moving) ions.
The real radiation
properties of a condensed magnetized surface can be intermediate between these limits
(see the discussion in \citealt{vAetal:05}).

In the free-ion approximation, there exist two broad absorption features in the spectrum.
The first one lies between the ion cyclotron energy
\be \label{u2_0}
E_{\rm c,i} = \hbar~ZeB/m_{\rm i}c \approx 0.0635~ B_{\rm 13}~ \left(\frac{Z}{A}\right)~~~\rm keV,
\ee
where $A$ is the mass number of the ion ($\approx$ 1 for H and $\approx$ 56
for Fe), $Z$ is the ion charge,  $m_{\rm i}$ is the ion mass,
$B_{\rm 13} = B/10^{13}$ G, and some boundary energy $E_{\rm C}$
\be \label{u2_m1}
E_{\rm C} \approx~ E_{\rm c,i} + E_{\rm p,e}^2/E_{\rm c,e},
\ee
 where  $E_{\rm p,e}$ is the electron plasma energy
\be \label{u2_m2}
     E_{\rm p,e} = \hbar \left(\frac{4\pi e^2 n_{\rm e}}{m_{\rm e}}\right)^{1/2} = 0.0288 ~
\left(\frac{Z}{A}\right)^{1/2} \left(\frac{\rho}{1~\rm g~ cm^{-3}}\right)^{1/2}~\mbox{keV},
\ee
and $E_{\rm c,e}$ is the electron cyclotron energy
\be \label{u2_m3}
E_{\rm c,e} = \hbar eB/m_{\rm e}c \approx 115.8~ B_{\rm 13}~~~\rm keV.
\ee
 Here $n_{\rm e}$ is the electron number density, $\rho$ is the mass
  density, and $m_{\rm e}$ is the electron mass. 

Only one radiation mode can
propagate
between $E_{\rm c,i}$ and  $E_{\rm C}$, therefore the radiation emissivity is reduced to approximately
1/2 in this range. The
condensed surface density depends on magnetic field and  chemical  composition (e.g., \citealt{Lai:01}):
\be \label{u2_1}
   \rho_{\rm s} \approx 8.9 \times 10^3~\eta'~ A~ Z^{~-3/5}~ B_{\rm 13}^{~6/5}~~~~ \rm g~cm^{-3},
\ee
where $\eta'$ is a coefficient ranging probably between 0.5 and 1 (see discussion in
\citealt{vAetal:05}).  Therefore,
the plasma  energy at the surface can be expressed  as \citep{vAetal:05}
\be \label{u2_2}
     E_{\rm p,e} \approx  2.7~\eta'^{1/2}~ Z^{1/5} B_{\rm 13}^{~3/5}~~~~\rm keV.
\ee
We conclude that it is possible to estimate the equivalent width of this absorption feature as:
\be \label{u2_3}
     EW \approx 0.5 \frac{E_{\rm p,e}^2}{E_{\rm c,e}} \approx
117~ \eta'~ (Z/26)^{2/5}~ B_{\rm 13}^{~1/5}~~~ \rm eV.
\ee
In all calculations below we used $\eta'=1$.

The second feature is at the plasma energy $E_{\rm pe}$, which is
not interesting for us. Indeed, the observed absorption lines lie
at $E <$ 1 keV, which would correspond to the plasma energy  at $B
< 6.4\times 10^{11}~~(Z/26)^{-1/3}$ G. Meanwhile, the
critical temperature for the plasma phase transition
leading to the formation of the condensed surface
is  roughly $ 7.5 \times 10^5 B_{\rm 13}^{~2/5}$ K
\citep{PCS:99,Lai:01}. Noting that the observed temperatures of the XDINSs
are
about $10^6$ K, we conclude that the absorption feature at the plasma energy in
the condensed surface radiation cannot be less than 1 keV in these stars.
Therefore, the observed absorption features in XDINSs
 cannot be associated with the absorption feature at the
plasma energy. However, the
first absorption feature (between $E_{\rm c,i}$ and $E_{\rm C}$) can
be prominent in their  spectra.

\begin{figure}
\begin{center}
\includegraphics[angle=0,scale=0.6]{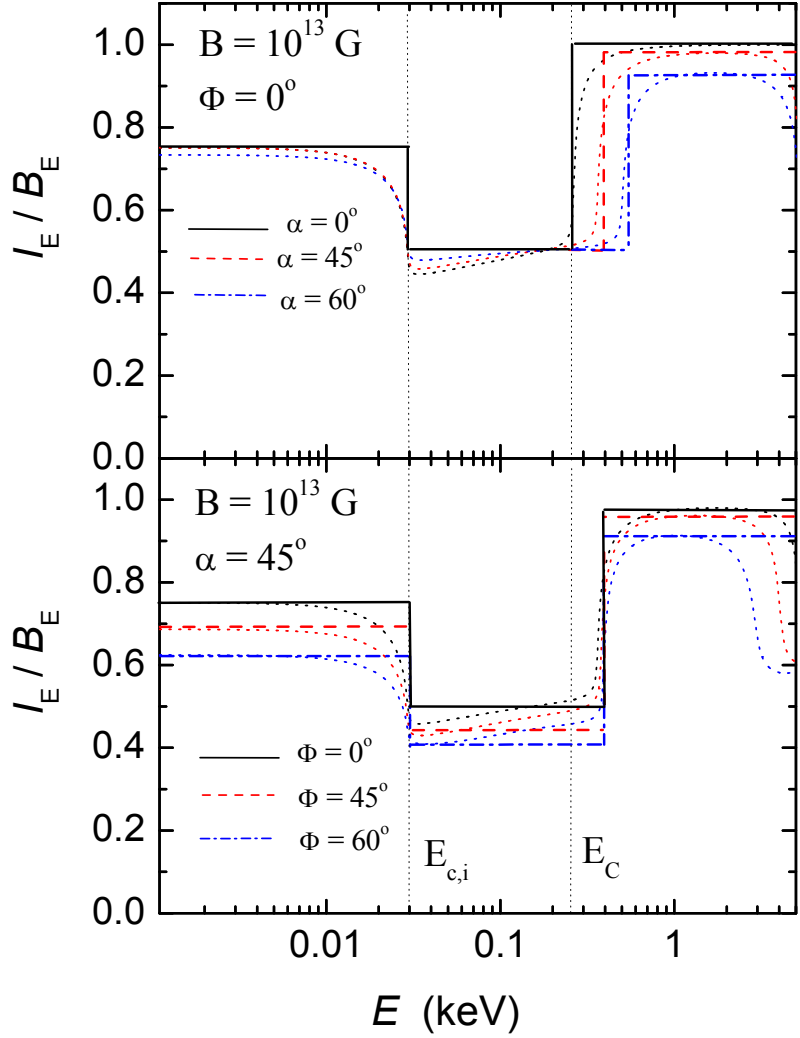}
\caption{\label{f:fig0}
Approximation of dimensionless emissivity as a function of photon energy $E$ for the case of
a condensed iron surface at $B = 10^{13}$ G. {\it Top panel:} The magnetic field is normal to surface.
The different curves correspond to different angles $\alpha$ between emergent photon
direction and surface normal. 
{\it Bottom panel:} The  emergent photon direction is inclined 45$\degr$
to the surface. The different curves correspond to different angles $\Phi$ between
magnetic field lines and surface normal.
 }
\end{center}
\end{figure}

The angular distribution of the condensed surface emissivity can be estimated from the  results of
 \cite{vAetal:05}:
\be \label{u2_4a}
   I_{\rm E} = (1-R)~B_{\rm E},
\ee
where $B_{\rm E}$ is Planck function,
\bea \label{u2_4}
   1-R & = & (1-0.27(1-\cos\alpha)^2)~(1-0.36(1-\cos\Phi)^2) \\ \nonumber
             & &~~~{\rm for}~~~E > E_{\rm C_1},
\eea
\be \label{u2_5}
   1 - R  =  0.3 + 0.2\cos\Phi
             ~~~{\rm for}~~~E_{\rm c,i} < E < E_{\rm C_1},
\ee
\be \label{u2_6}
   1-R  = (0.5+0.25\cos\Phi)
             ~~~{\rm for}~~~E < E_{\rm c,i},
\ee
and
\be \label{u2_7}
   E_{\rm C_1} = E_{\rm C}~(1+3(1-\cos\alpha)^{3/2}).   
\ee
Here $\alpha$ is the angle between radiation propagation direction and the
surface normal, and $\Phi$ is the angle between the surface normal and the
magnetic field lines. This approximation is presented graphically in
Fig.~\ref{f:fig0} together with accurate calculations, performed by the method described by
\cite{vAetal:05} in the free ions approximation.

 Here we neglect the angular dependence on the
azimuthal angle $\varphi$ between radiation propagation plane and the
normal-magnetic field line plane for the inclined magnetic field (see Fig.~6 in
\citealt{vAetal:05}) and the absorption feature at $E_{\rm pe}$.  We also ignore
the damping effects, which smooth the transitions between different bands.  
That approximation leads to sharp boundaries at $E_{\rm c,i}$ and $E_{\rm C_1}$
in the computed spectra (Figs.\, \ref{f:fig3}, \ref{f:fig4}, \ref{f:fig4a} -
\ref{f:fig6}, \ref{f:fig9}), which can be smoother in more detailed computations
(see, however, next paragraph).  In our approximation we ignore the azimuthal
dependence of $R$ in the case of an inclined magnetic field.  At some azimuthal
angles the considered absorption feature can be less significant (see Fig.\,4 in
\citealt{vAetal:05}).  Therefore, our calculations with this approximation will
give an upper limit for the strength of the absorption feature.  This effect
must be less significant in the case of a thin model atmosphere above the
condensed surface due to significant atmospheric optical depth in the X-mode
around the proton cyclotron resonance (see Fig.\,\ref{f:tau}).

Our approximation appears rough, but the more accurate calculations of $R$ have
significant uncertainties, too. First of all,  these calculations were performed
in two different approximations: a complete neglect of ion motions,  and free
ions. The real picture is in some sense intermediate. Ions in the crystal grid
(or in liquid) are not free, but can move.  These two limits give very different
spectra at energies below $E_{\rm c,i}$ (see Figs. 2--6 in \citealt{vAetal:05}).  
The approximately known factor $\eta'$, which
determines the value of $E_{\rm C}$ and the width of the absorption feature, was
already mentioned above. Moreover, the accurate calculations are only applicable
for a perfectly smooth condensed surface. If the surface has some roughness, the
emitted spectrum  has to be close to a blackbody \citep{vAetal:05}. Our
approximation corresponds to the free-ions limit and gives  radiation spectra,
that reflect the main features of condensed surface spectra, especially if we
take into account all uncertainties mentioned above.


\subsection{Model atmospheres of magnetized neutron stars}

Model atmospheres for magnetized neutron stars were computed by many scientific groups, but we
use our own calculations for the description of their radiation spectra. We employ the code
presented by \cite{sul:09}, where details of the computational method can be found.

Here we assume that the magnetic field strengths of the XDINSs are about a few $\times 10^{13}$ G and
consider a pure hydrogen atmosphere. The effective temperatures are $\approx 10^6$ K.
At these values of $B$ and $T$
hydrogen is partially ionized \citep{PCS:99}
and the vacuum polarization effect together with the
partial mode conversion can be significant \citep{Ho.Lai:03,vAL:06}.
Therefore the model atmospheres presented below include these effects.
 The magnetic field in presented models is assumed to be
homogeneous and normal to the surface.
In general, it is necessary to study model atmospheres with inclined magnetic field
for the modeling of integral
spectra of neutron stars \citep{Hoetal:08}. But here we
consider general features
of the model spectra, and therefore we neglect
the distinctions between spectra of the models with different
field inclinations, which are unimportant for the
qualitative behavior of the spectrum  (see corresponding figures in \citealt{Shibanovetal:92,sul:09}).
Besides, the bright regions at magnetic poles of a star in our models considered below
have magnetic fields close to normal.

\begin{figure}
\begin{center}
\includegraphics[angle=0,scale=0.6]{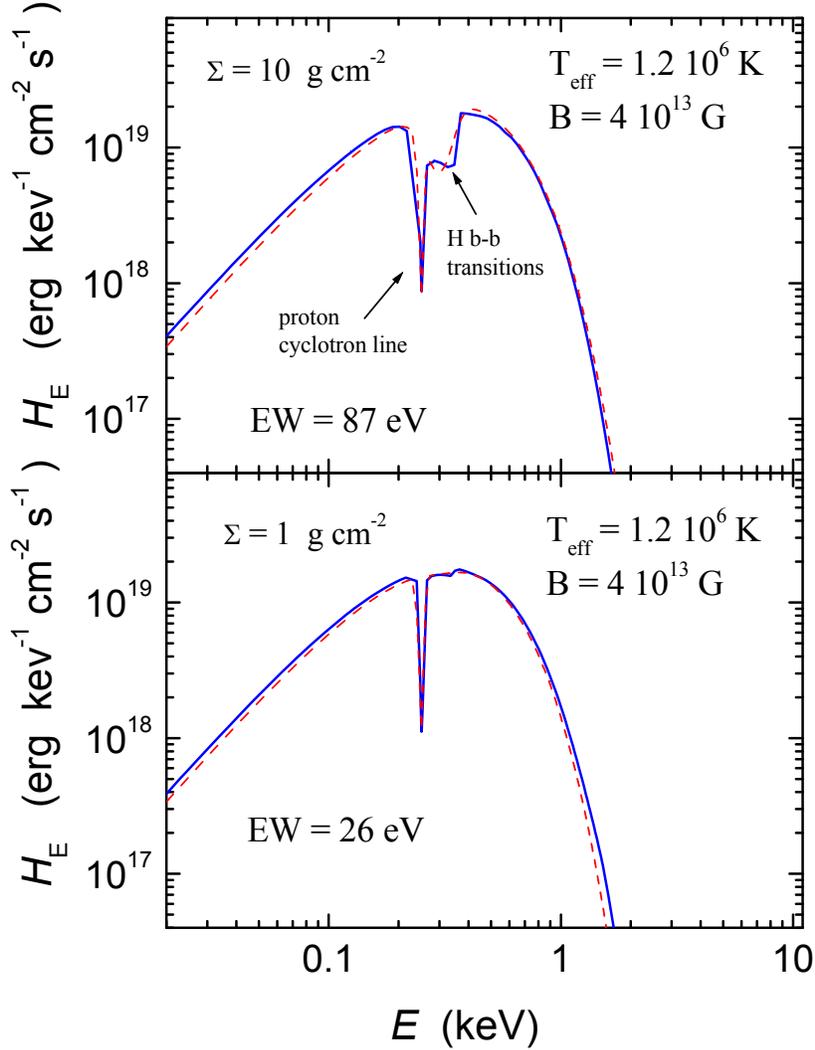}
\caption{\label{f:fig2}
Emergent spectra of two thin model atmospheres above a blackbody
condensed surface (solid curves) with $T_{\rm eff} = 1.2 \times 10^6$ K and
 $B = 4 \times 10^{13}$ G together with fitting spectra (dashed curves). The fitting spectra are
blackbody spectra with the color temperature $T_{\rm c} = f_{\rm c} T_{\rm eff}$ and two Gaussian lines.
{\it Top panel:} The spectrum of the model with surface density $\Sigma$ = 10 g cm$^{-2}$.
Parameters of the fitting  spectrum: $f_{\rm c} = 1.2$, $D = 1.14^{-4}$,  center of the first line (proton cyclotron) $E_1 = 0.25$ keV,
 optical depth  $\tau_1$ = 2.6, and width $\sigma_1$ = 6 eV,
 and for the second line (H bound-bound transitions) $E_2 = 0.3$ keV,  $\tau_2$ = 1.1,
 $\sigma_2$ = 40 eV. {\it Bottom panel:} The spectrum of the model
with  surface density $\Sigma$ = 1 g cm$^{-2}$. Parameters of the fitting 
 spectrum: $f_{\rm c} = 1.12$, $D = 1.12^{-4}$,  line parameters are the same
as above, except $\tau_2$ = 0.08.
 }
\end{center}
\end{figure}

The emergent spectra of two thin magnetized
model atmospheres (i.e. a hydrogen layer above a condensed surface) with 
parameters typical for the XDINSs are shown in Fig.~\ref{f:fig2}.
 We display the Eddington flux
\be \label{edflx}
   H_{\rm E} = \frac{1}{2} \int_0^{\pi/2} I_{\rm E}(\cos\alpha) \cos\alpha \sin\alpha\,d\alpha,
\ee
where $I_{\rm E}$ is a specific intensity. Later we also use the astrophysical flux $F_{\rm E}=4 H_{\rm E}$.
The often used physical flux is defined as $\cal{F}$$_{\rm E}=\pi F_{\rm E}=4\pi H_{\rm E}$. 
For blackbody radiation it is  $F_{\rm E}= I_{\rm E}= B_{\rm E}$.
In the models shown in Fig.2 the  blackbody radiation is used as the inner boundary condition 
for the radiation transfer equation.
There are absorption features in the spectra, which arise due to the proton cyclotron
absorption and bound-bound and bound-free transitions in hydrogen atoms. 
 It is necessary to remark that the proton
cyclotron feature is weakened by vacuum polarization and  partial mode conversion, but
nevertheless remains prominent in the spectra, as can be seen in our figures below
(see also \citealt{sul:09}).
Equivalent widths (EWs) 
of these features are $\approx$ 25 eV and  $\approx$ 90 eV
for the models with surface densities $\Sigma$ = 1 and 10 g cm$^{-2}$,
respectively.  The absorption 
feature in the spectrum of a
semi-infinite atmosphere is even less prominent in comparison to the absorption feature in the model
with $\Sigma$ = 10 g cm$^{-2}$ (see \citealt{sul:09}) and its absorption feature
has a smaller EW ($\approx$ 50 eV). The EW of the
absorption feature slightly depends on the magnetic field strength and the relation between proton
cyclotron energy and $kT_{\rm eff}$. For example, the EW of the absorption feature in the spectrum of the model
with $B = 10^{14}$ G and $\Sigma$ = 10 g cm$^{-2}$ and the same $T_{\rm eff}$ is close to 80 eV
(see Fig.~\ref{f:fig4}).

The spectra of these models can be fitted approximately by diluted blackbody spectra $B_{\rm E}$ with
a color temperature $T_{\rm c}$ slightly larger than the effective temperature $T_{\rm eff}$ and two
Gaussian absorptions:
\be
\label{u2_8}
     F_{\rm E} = D~B_{\rm E}(f_c~T_{\rm eff})~ \exp(-\tau_1)~ \exp(-\tau_2),
\ee
where $f_c = T_{\rm c}/T_{\rm eff}$ is the hardness factor, $D$ is the dilution factor, which can differ
from the  usual $f_c^{\,-4}$.  Later we will present the dilution factor in the
  form $D = f_c^{\,'\,-4}$  to distinguish it from the usual value. 
The optical depths of the absorptions at a given photon energy $E$ are
\be
\label{u2_9}
      \tau_{1,2} = \tau_{1,2}^{0} \exp\left(-\frac{(E-E_{1,2})^2}{2\sigma_{1,2}
^2}\right).
\ee
Here $\tau_{1,2}^{0}$ are  the optical depths at the centers of the lines at energy $E_{1,2}$ and $\sigma_{1,2}$
are the line widths. The values of these fitting parameters for these two spectra are given
in the caption of Fig.~\ref{f:fig2}.  The first line corresponds to the proton cyclotron line in the model atmosphere spectra,
and the second one corresponds to the bound-bound transitions in
hydrogen atoms (cf.\ \citealt{PCh:04,Potekhinetal:04}).

It is clear that the blackbody radiation model
is not an accurate inner boundary condition for the thin model
atmospheres above the condensed surfaces, and the real radiation properties
of the surface have to be used for
precise modeling.
Below we use our approximations (\ref{u2_4}) -- (\ref{u2_6}) for the inner boundary conditions of radiation
transfer equation for both radiation modes, extraordinary (X) and ordinary (O):
\be \label{u2_10}
        I^{\rm ~X}_{\tau_{\rm max}}(\eta,E) = \frac{1}{2}B_E(T(\tau_{\rm max})) (1-2R) +
2R~I^{\rm ~X}_{\tau_{\rm max}}(-\eta,E),
\ee
\be \label{u2_11}
        I^{\rm ~O}_{\tau_{\rm max}}(\eta,E) = \frac{1}{2} B_E(T(\tau_{\rm max})).
\ee
where $\eta = \cos \alpha$, and $I^{\rm~X,O}$ are the specific intensities of the extraordinary and ordinary modes.
In future work we plan to replace this approximation
by accurately computed radiation properties of
the condensed surfaces. But we expect that
the present approximation is qualitatively correct.

\begin{figure}
\begin{center}
\includegraphics[angle=0,scale=0.4]{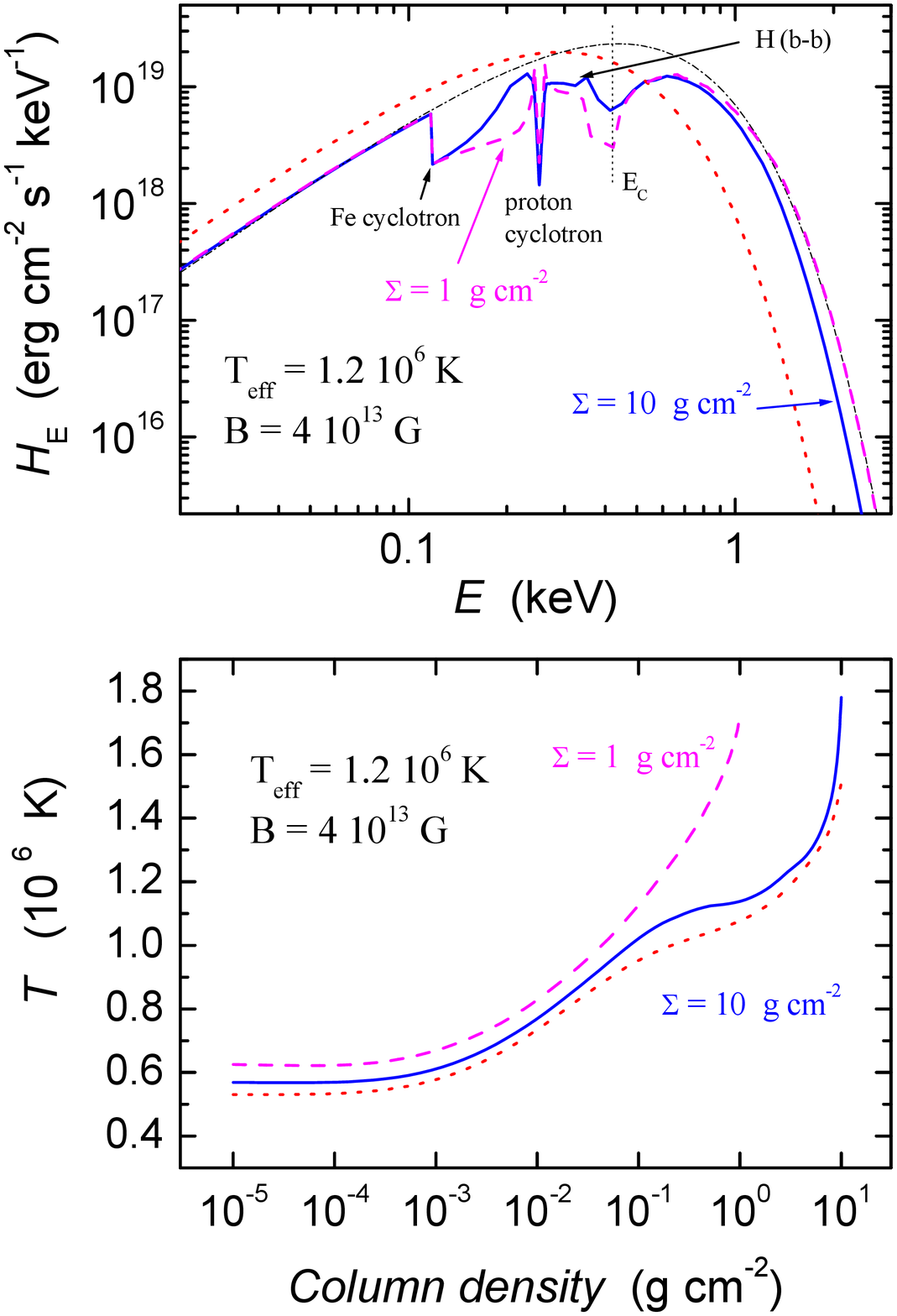}
\caption{\label{f:fig3}
Emergent spectra (top panel) and temperature structures (bottom panel) of thin
partially ionized hydrogen atmospheres ($T_{\rm eff} = 1.2 \times 10^6$ K,
$B = 4 \times 10^{13}$ G) above a condensed iron surface with
$\Sigma$ = 10 g cm$^{-2}$ (solid curves) and $\Sigma$ = 1 g cm$^{-2}$ (dashed curves).
An inner boundary condition for the radiation transfer equation, corresponding to the condensed
surface radiation properties (see text), was used. The blackbody spectra
with $T = T_{\rm eff}$ (dotted curve) and the diluted blackbody  spectra
($f_{c}$ = 1.5, $D = 1.3^{-4}$), which describe the
high-energy part of the spectrum of the atmosphere with $\Sigma$ = 1 g cm$^{-2}$ (dash-dotted curve)
are shown for comparison in the top panel. In the bottom panel the temperature structure of the model with
$\Sigma$ = 10 g cm$^{-2}$ using blackbody radiation as the inner boundary condition is also shown
(dotted curve).
 }
\end{center}
\end{figure}

\begin{figure}
\begin{center}
\includegraphics[angle=0,scale=0.4]{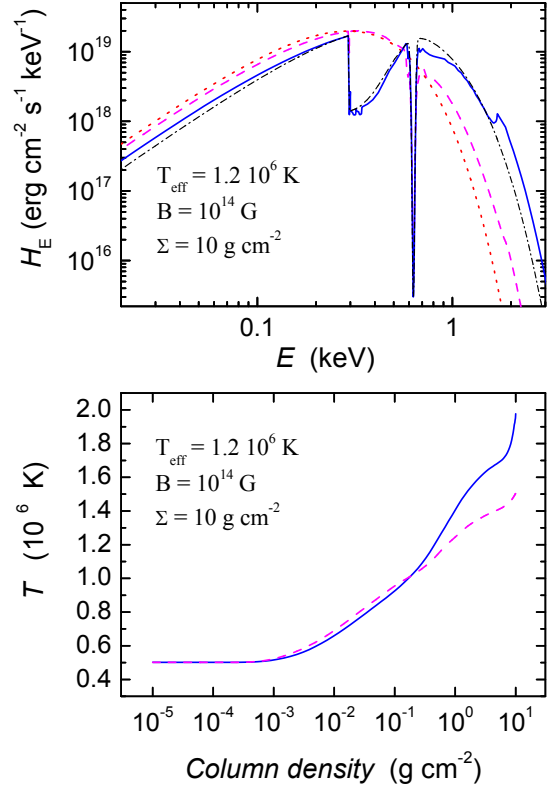}
\caption{\label{f:fig4}
Effect of the inner boundary condition of the radiation transfer equation on
emergent spectrum (top panel) and temperature structure (bottom panel) of a thin
partially ionized hydrogen atmosphere ($T_{\rm eff} = 1.2 \times 10^6$ K,
$B =  10^{14}$ G) with $\Sigma$ = 10 g cm$^{-2}$, above the condensed iron surface (solid curves)
and blackbody surface (dashed curves). For comparison the blackbody spectrum
with $T = T_{\rm eff}$ (dotted curve) and the diluted blackbody  spectrum
  ($f_c$ = 1.6, $D = 1.4^{-4}$),   with a Gaussian line representing the proton cyclotron line ($E_1 = 0.63$ keV,
$\tau_1$ = 8.5, $\sigma_1$ = 11 eV), and a half-Gaussian line ($E_2 = 0.3$ keV,
$\tau_1$ = 2.5, $\sigma_1$ = 150 eV) describing the absorption feature at energies higher than $E_{\rm c,i}$ (dash-dotted curve)
are shown.
That last curve fits the spectrum of the atmosphere above the condensed iron surface.
}
\end{center}
\end{figure}

\begin{figure}
\begin{center}
\includegraphics[angle=0,scale=0.6]{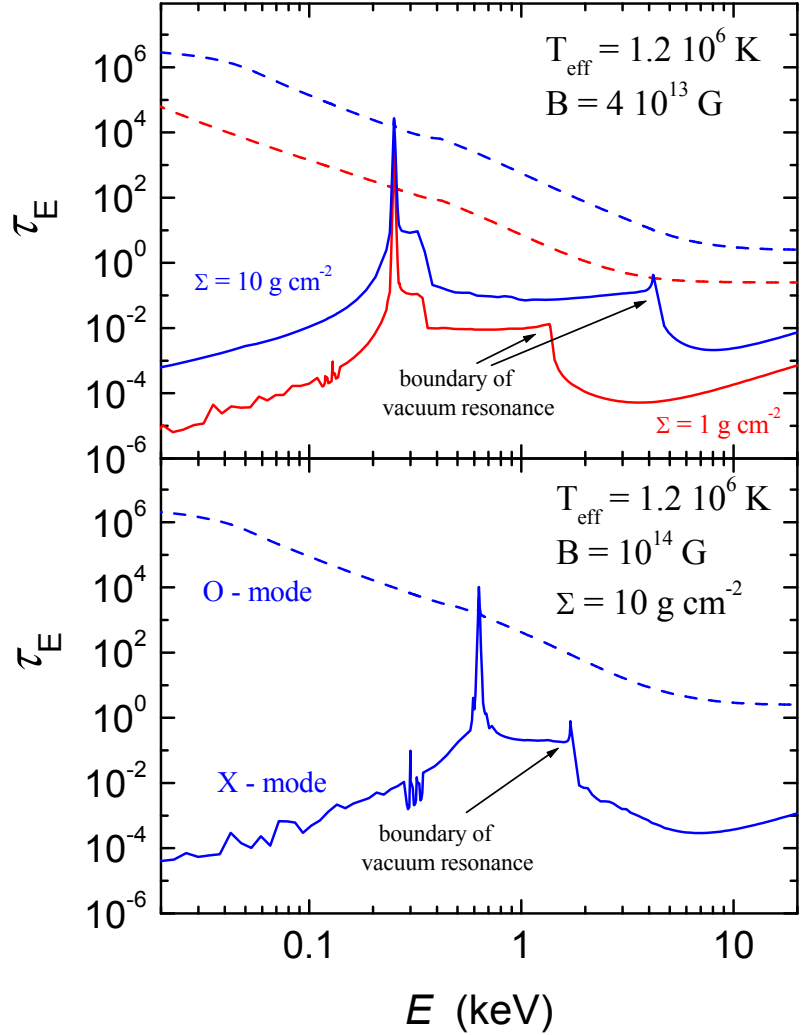}
\caption{\label{f:tau}
 Optical thickness of a thin
partially ionized hydrogen atmospheres with $T_{\rm eff} = 1.2 \times 10^6$ K
 above the condensed iron surface in X-mode (solid curves)
and O-mode (dashed curves). {\it Top panel:} $B =
4\times 10^{13}$ G and $\Sigma$ = 1 and 10 g cm$^{-2}$.
{\it Bottom panel:} $B =  10^{14}$ G and $\Sigma$ = 10 g cm$^{-2}$. 
}
\end{center}
\end{figure}

\begin{figure}
\begin{center}
\includegraphics[angle=0,scale=0.6]{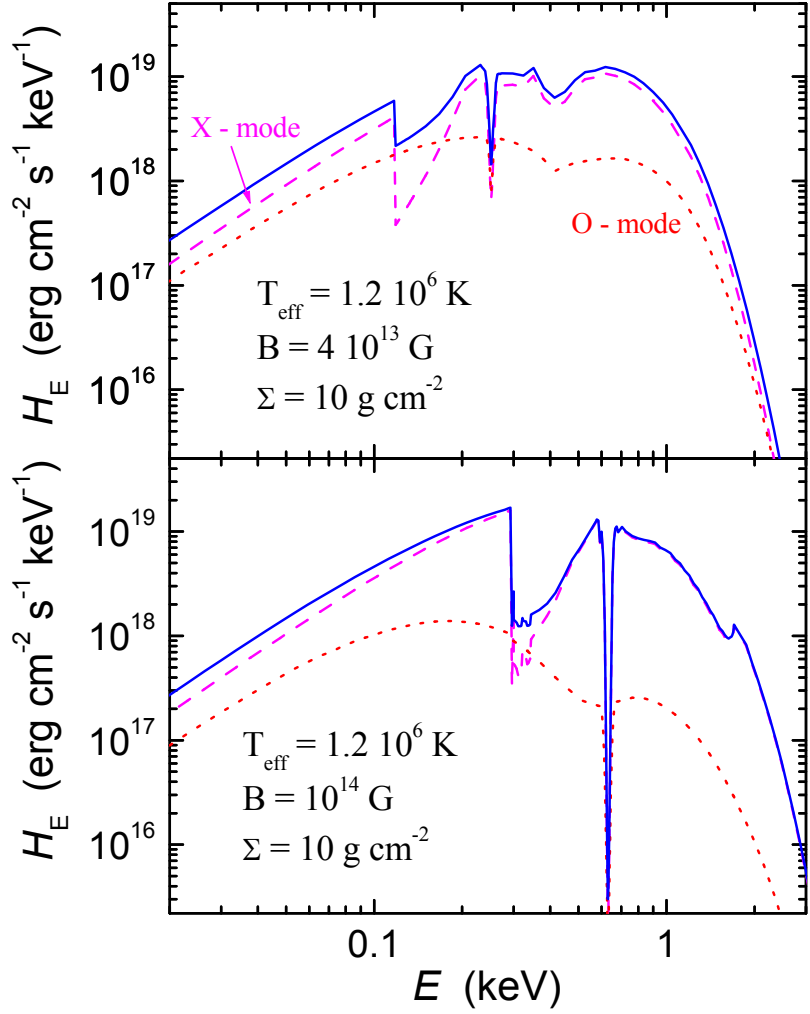}
\caption{\label{f:fig4a}
Total emergent spectra  (solid curves) together with spectra in
the X-mode (dashed curves) and in the O-mode
(dotted curves) of a thin
partially ionized hydrogen atmosphere  with $T_{\rm eff} = 1.2 \times 10^6$ K,
 $\Sigma$ = 10 g cm$^{-2}$ above the condensed iron surface  with  $B =
4\times 10^{13}$ G (top panel), and $B =  10^{14}$ G (bottom panel).
 }
\end{center}
\end{figure}

\begin{figure}
\begin{center}
\includegraphics[angle=0,scale=0.6]{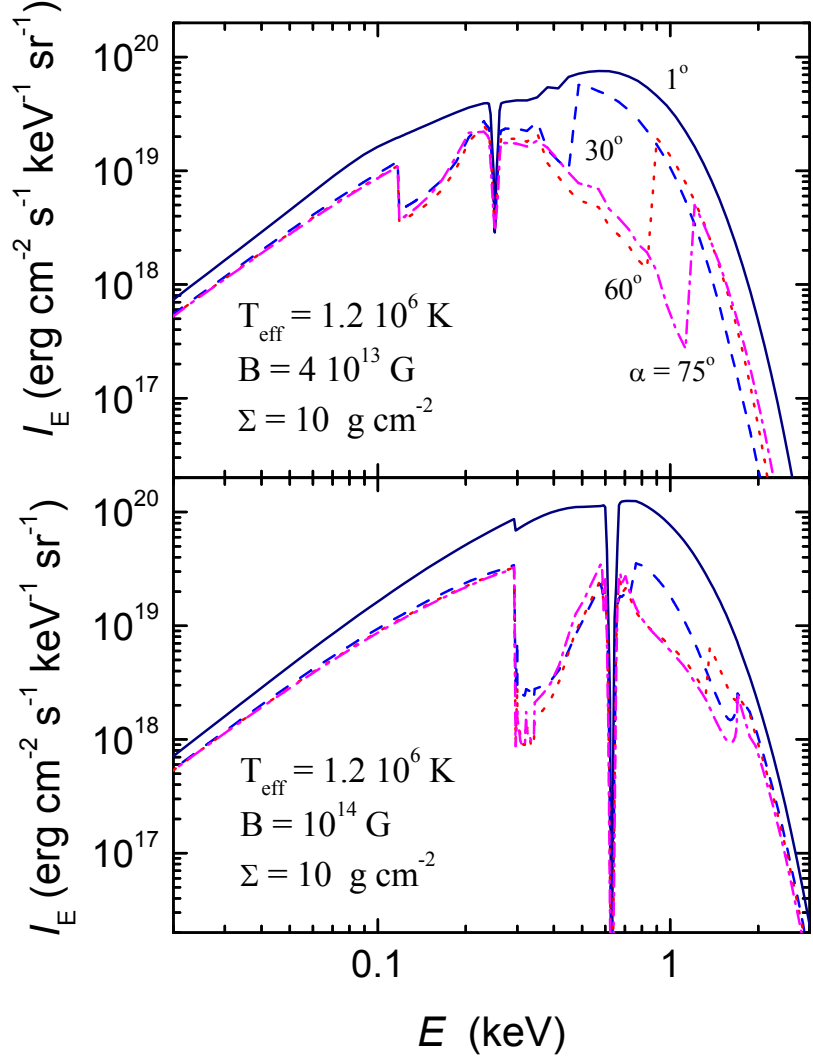}
\caption{\label{f:fig5}
Emergent specific intensity spectra of two  model atmospheres above a condensed iron surface
 with $T_{\rm eff} = 1.2 \times 10^6$ K, $\Sigma$ = 10 g cm$^{-2}$,  and
$B = 4 \times 10^{13}$ G (top panel) and $B =  10^{14}$ G (bottom panel). Spectra at different angles
between radiation propagation and surface normal are shown: 1${\degr}$ (solid curves),
30${\degr}$ (dashed curves), 60${\degr}$ (dotted curves) and  75${\degr}$ (dash-dotted curves).
 }
\end{center}
\end{figure}

\begin{figure}
\begin{center}
\includegraphics[angle=0,scale=0.6]{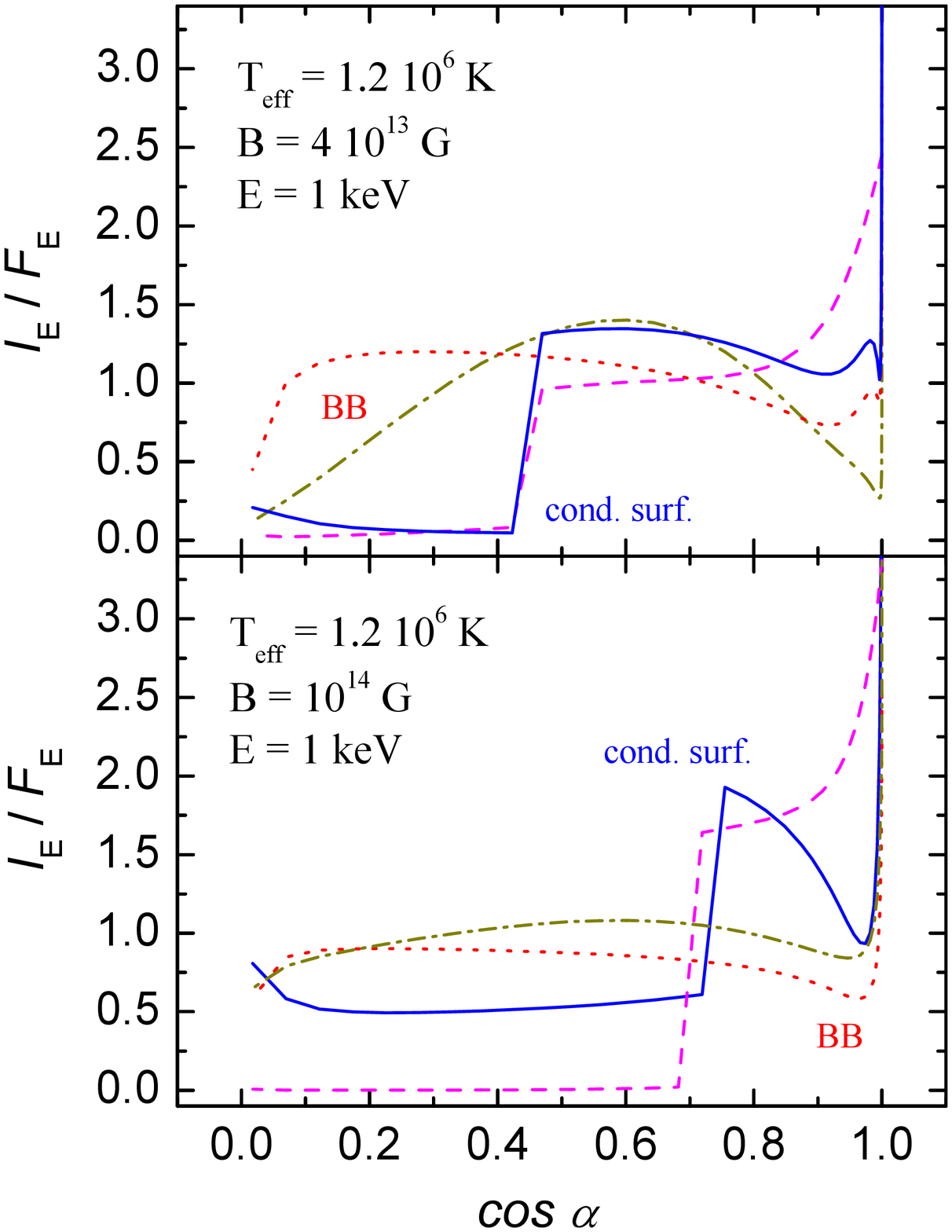}
\caption{\label{f:fig6}
Angular distributions of the emergent radiation at photon energy 1 keV.
{\it Top panel:} Models with $T_{\rm eff} = 1.2 \times 10^6$, $B =  4 \times 10^{13}$ G
above the condensed iron surface and $\Sigma$ = 10 g cm$^{-2}$ (solid curve), $\Sigma$ = 1 g cm$^{-2}$
(dashed curve), $\Sigma$ = 10 g cm$^{-2}$ above a blackbody (dotted curve), and a semi-infinite
atmosphere (dash-dotted curve).
{\it Bottom panel:} Models with $T_{\rm eff} = 1.2 \times 10^6$, $B =  10^{14}$ G
 and  above the condensed iron surface with $\Sigma$ = 10 g
  cm$^{-2}$ (solid curve), $\Sigma$ = 1 g cm$^{-2}$ (dashed curve), $\Sigma$
  = 10 g cm$^{-2}$ above a blackbody (dotted curve), and a semi-infinite
atmosphere (dash-dotted curve).
 }
\end{center}
\end{figure}

We have
computed  thin hydrogen model atmospheres above condensed iron surfaces
using these new inner boundary conditions. The results are presented in Figs.~\ref{f:fig3} --
\ref{f:fig6}. Model atmospheres  are optically thick in the O-mode in the observed part of
the X-ray band (0.1 -- 5 keV), and almost everywhere optically thin
in the X-mode. Therefore, the total spectra
are defined mostly by the X-mode spectra, because the O-mode
spectrum arises in the cooler layers of the atmosphere.

The emergent spectra of the thin magnetized model atmospheres above the condensed surface are harder
($f_c \sim$ 1.5 - 1.7) in comparison to the models above a blackbody, because
the temperatures at the inner boundaries of these models are higher 
(see bottom panels in Figs.~\ref{f:fig3}
-- \ref{f:fig4}).

Due to the condensed surface emission properties, there
appear broad absorption features between  $E_{\rm c,i}(\rm Fe)$ and
approximately $E_{\rm C}$, especially prominent at $E_{\rm c,i}$.
 The dependence of optical depths of thin atmospheres in both modes
  on the photon energy are shown in Fig.~\ref{f:tau}. The atmospheres are
  optically thin in X-mode at almost all energies, except the region around
  the proton cyclotron energy. The total spectra are dominated by the X-mode (see
  Fig.~\ref{f:fig4a}), therefore, they are close to the condensed iron spectra
  in X-mode, especially in the Wien tail. But this is not the case for the
  band between $E_{\rm c,i}(\rm Fe)$ and $E_{\rm C}$. 
In this region the O-mode flux can contribute a  significant part to the total flux
(see Fig.~\ref{f:fig4a}). At energies in the vicinity of $E_{\rm
c,i}(\rm H)$ the opacity in the X-mode increases, therefore, the optical depth of atmosphere 
 and the emergent flux increase, too. As a result a
complex absorption feature arises with a total EW $\approx$ 400 eV
for the model with $B = 10^{14}$ G
 and $\Sigma$ = 10 g cm$^{-2}$,
  EW $\approx$ 370 eV for 
 $B = 4 \times 10^{13}$ G
 and  $\Sigma$ = 1 g cm$^{-2}$,
  and EW $\approx$ 260 eV for
 $B = 4 \times 10^{13}$ G
 and $\Sigma$ = 10 g cm$^{-2}$. It is interesting to
remark that the EWs of these features are larger than the sums of
the EWs
of the cyclotron line in the atmosphere spectra and the EWs of
the absorption feature in the condensed surface spectra (see
Eq.\,(\ref{u2_3})). The reason is  the following. Between $E_{\rm
c,i}$ and  $E_{\rm C}$ a naked condensed surface radiates  mainly
in the O-mode, which is approximately half of the flux in comparison to other
energies. In contrast, a thin atmosphere above it is optically thick in
the O-mode, and the emergent flux forms in the upper atmosphere layers,
which have significantly lower temperatures than the
condensed surface. Therefore, the emergent flux in this band is
significantly smaller compared to the  naked condensed
surface. This is particularly noticeable in the model with $B =
10^{14}$ G.

Apart from the spectral energy distribution of the emergent flux,
the angular distribution is also important. Examples 
 for various  models of magnetized atmospheres
are shown in Figs.~\ref{f:fig5} -- \ref{f:fig6}. There is
 a narrow spike at the
surface normal, but the total energy radiated in this spike is
relatively small and we don't consider these spikes below. Angular
distributions of semi-infinite magnetized atmospheres have
maxima at $\alpha \approx$ 40${\degr}$ -- 60${\degr}$
\citep{pavlovetal:94}. An example  of such a distribution is
presented in the bottom panel of Fig.~\ref{f:fig6}. It
can be fitted by the function
\be
\label{u2_12}
   I_E (\alpha) = F_E \times \phi(\alpha),
\ee
where
\be
\label{u2_13}
   \phi(\alpha) \approx 3.6 \cos\alpha - 0.5 \cos^2\alpha - 2.9 \cos^3\alpha.
\ee

The angular distributions of the specific intensities of the
thin magnetized atmospheres above a  blackbody are close to
isotropic (Fig.~\ref{f:fig6}). The corresponding
distributions for the atmospheres above condensed surfaces are
also almost isotropic ones (see Fig.\ref{f:fig6}), but  have some
interesting feature at energies between  $E_{\rm C}$ and 4$E_{\rm
C}$. The location of the high energy boundary of
the broad  depression depends on $\alpha$ (see Eq.\,(\ref{u2_7})).
Therefore, at a given photon energy there is an angle $\alpha$ 
corresponding to this location. At this angle the specific intensity
angular distribution has a jump. At smaller angles the distribution
 corresponds to the usual angular distribution for the model
atmosphere, but at the larger angles the specific intensities are
significantly smaller (see Fig.~\ref{f:fig6}). This jump is more
prominent for the model with $B = 4 \times 10^{13}$ G and it is also
well visible in the spectra of the specific intensities for this
model (the top panel of Fig.~\ref{f:fig5}). In the angular
distributions of the intensities of the  model with $B = 10^{14}$
G this jump is less significant (Figs.~\ref{f:fig5} and
\ref{f:fig6}), because  for this model $E_{\rm C} \approx$
$E_{\rm c,i}(H)$ and the X-mode opacity is more important, together
with the mode conversion. The features of the magnetized model
atmospheres are very unusual and they have to be studied in more
detail with a more accurate treatment of the radiation properties of the
condensed surfaces.    

 From the presented examples of the partially ionized hydrogen magnetized model atmospheres we can conclude,
that they display complex absorption features with EW $\la 100$ eV in the spectra.
The naked condensed surfaces can have broad absorption features with slightly larger EWs ($\ge$ 120 eV), but
the absorption structures in the spectra of thin atmospheres above condensed surfaces can have
significantly larger EWs, up to 400 eV.

\section {Calculation method for the integral neutron star spectrum}
\label{s:imodel}

We consider a slowly rotating spherical isolated neutron star with given compactness $M/R$ as a
model for XDINSs spectra and light curves fitting, taking into
account distributions of the local
temperatures and magnetic field strengths over the stellar surface. Local
spectra are blackbody-like 
spectra with corresponding temperature and
absorption lines, whose central energies depend on the local magnetic field. Gravitational redshift
and the light bending are taken into account.

\begin{figure}
\begin{center}
\includegraphics[angle=0,scale=0.6]{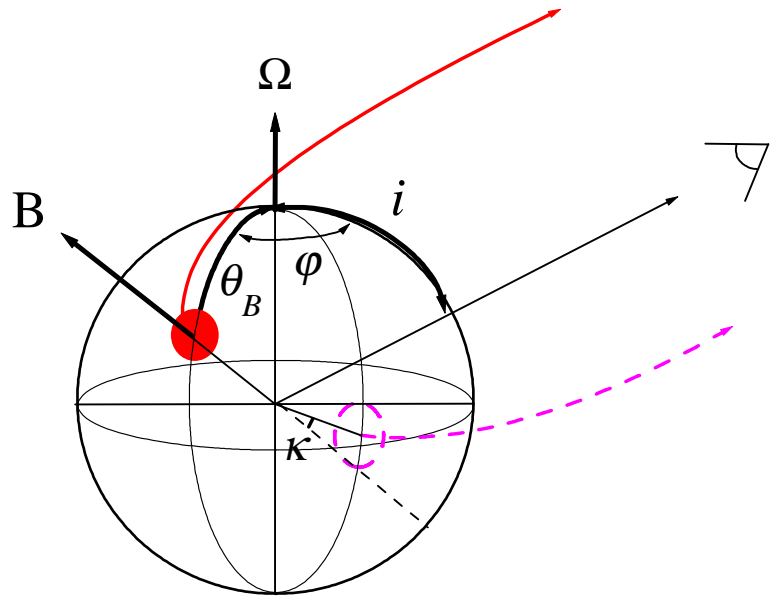}
\caption{\label{f:fig1}
Geometry of the model. The opposite bright spot is indicated by the dashed circle.
 }
\end{center}
\end{figure}

The geometry of our model is shown in Fig.~\ref{f:fig1}. The magnetic axis is inclined to the rotation
axis by the angle $\theta_{\rm B}$, and $i$  is the inclination angle of the rotation axis relative to the line of sight.
The phase angle $\varphi$ is the angle between the plane containing
the rotation axis and the line of sight and the plane containing
the magnetic and rotation axes.
One magnetic pole can be shifted relative to the antipode
of the other pole by
a small angle $\kappa$. We assume that
the brightest areas of the NS surface are located at the magnetic poles. Therefore,
temperature and magnetic field distributions near the poles are
most important. 

XDINSs pulse fractions are significant, and a classical temperature
distribution, which arises due to a global (core) dipole  magnetic
field of a NS, is not sufficient to reproduce observed pulse fractions
\citep{Page:95,Geppertal:06}. \cite{Swopeetal:05} showed that
the RBS\,1223 light curves can be fitted  by two models,
namely two bright isothermal spots with different temperatures, or two
narrow peaked temperature distributions  (with different
parameters) taken from \cite{Geppertal:04}. \cite{Geppertal:04}
considered the thermal  transport in the NS crust for a poloidal
dipole magnetic field, concentrated in the crust only.  In this
case the magnetic field lines have large inclinations to the
surface normal in a wider band around the magnetic
equator than in the global dipole field case. The thermal
conductivity across magnetic field lines is smaller than along them,
therefore the bright spots around  magnetic poles are brighter
and smaller than they would be for the dipole distribution.
Another possibility is a strong toroidal magnetic
field in the NS crust. Such kind of models were computed by
\cite{PerAzetal:06a} using various approximations. They obtained
an analytical approximation of the surface temperature
distribution for the crust magnetic field with a strong toroidal component
calculated for the force-free conditions. We use this approximation for
the temperature distribution in our model.

\begin{figure}
\begin{center}
\includegraphics[angle=0,scale=0.6]{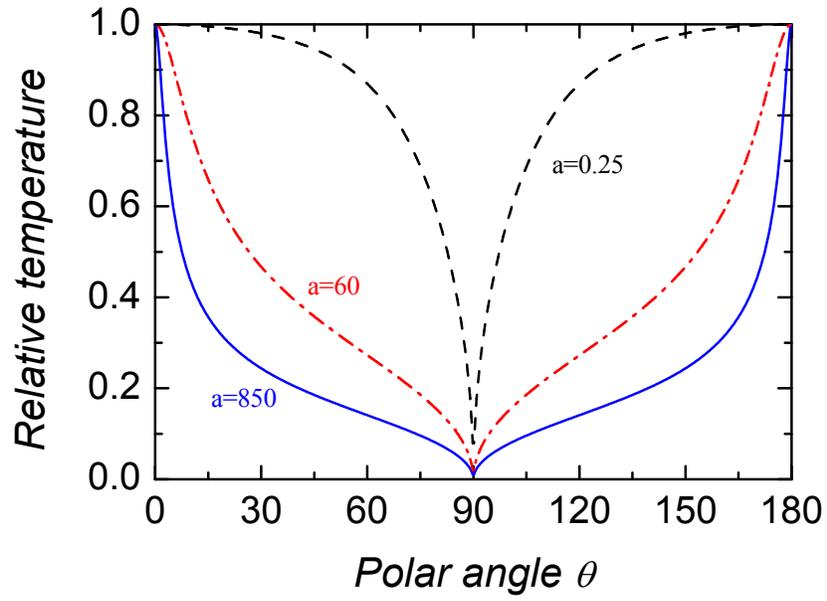}
\caption{\label{f:fig2a}
Temperature distributions at various parameters $a$.
Only the first term on the right-hand side of Eq.\,(\ref{um4})
is used for these calculations.
 }
\end{center}
\end{figure}

In a well known approximation by \citet{Gren.Hart:83} the temperature distribution is:
\be
\label{um1}
  T^4 = T_{\rm p}^4 \left(\cos^2\Phi+\frac{K_{\bot}}{K_{\parallel}}\sin^2\Phi\right),
\ee    
 where  $K_{\bot}$  and  $K_{\parallel}$ are thermal conductivities across and along
the magnetic field, respectively, and
$T_{\rm p}$ is the temperature at the magnetic pole. Magnetic fields of the isolated NSs can be
 strong enough ($\sim 10^{13}$ G) so that the ratio $K_{\bot}/{K_\parallel}$ is small. In this case the
temperature distribution can be approximated as \citep{PerAzetal:06b}:

\be
\label{um2}
  T^4 = T_{\rm p}^4 \cos^2\Phi + T_{\rm min}^4.
\ee    
In the force-free approximation a simple relation between the magnetic colatitude $\theta$ and the magnetic field
inclination in the NS crust was suggested by
\citet{PerAzetal:06a}:
\be
\label{um3}
  \cos^2\Phi = \frac{4\cos^2\theta}{(1+\mu^2R^2)+(3-\mu^2R^2)\cos^2\theta},
\ee    
where $\mu^{-1}$ is a typical toroidal magnetic field length scale, and $\mu R$ is an approximate  ratio
of the toroidal component of the magnetic field to the poloidal one. The case $\mu$=0
corresponds to a core dipole magnetic field.
 Although this magnetic field model was derived for the crust,
we use Eq.(\ref{um3})  also in the atmosphere, because it allows us to
investigate the possible effects of magnetic field and temperature
inhomogeneity by changing a single parameter $\mu$.
Finally, we use the following
temperature distributions around magnetic poles:
\be
\label{um4}
  T^4 = T_{\rm p1,2}^4 \frac{\cos^2\theta}{\cos^2\theta + a_{\rm 1,2}\sin^2\theta } + T_{\rm min}^4,
\ee    
where  $a_{\rm 1,2} = (1+\mu_{\rm 1,2}^2R^2)/4$ and $T_{\rm p1,2}$ are the temperature distribution
parameters. This can describe various temperature distributions
(see Fig.~\ref{f:fig2a}), from strongly
peaked ($a \gg 1$) to the classical dipolar ($a = 1/4$) and homogeneous
($a = 0$).

Another possibility is to consider two isothermal bright spots around the magnetic poles with fixed temperatures and
angular sizes.

The surface magnetic field distribution, which corresponds to
the magnetic field inclination  in the force-free approximation
model (Eq.\,\ref{um3}) is described as follows:
\be
\label{um5}
B = B_{\rm p1,2} \sqrt{\cos^2\theta + a_{1,2}\sin^2\theta},
\ee
where $B_{\rm p1,2}$ are the magnetic field strengths at the poles.

Temperature and magnetic field distributions are calculated separately around each
 magnetic pole, therefore, the distributions are not smooth at the
magnetic equator area if one of the poles is shifted.
But this small mismatch at the equator is unimportant,
because this area is dim.  

The local spectrum of the NS surface in the case of magnetized atmospheres at given temperature and
 magnetic field is taken as a diluted blackbody
spectrum with two additional absorption features
(see also Eqs.\,(\ref{u2_8}), (\ref{u2_9}) and (\ref{u2_12}))
\be
\label{um7}
     I_{\rm E}(\alpha) = D~ B_{\rm E}(f_c T)~\phi(\alpha)\,\exp(-\tau_1)\,\exp(-\tau_2),
\ee
where the optical depth $\tau_i$ of the absorption line $i$ at given photon energy $E$ is
defined by Eq.\,(\ref{u2_9}), and
$\tau_{\rm 1,2}^{\rm 0}$ , $\sigma_{\rm 1,2}$ are considered as  model parameters identical 
at all NS surface elements.
The photon energy of the line center $E_i$ depends on the
local magnetic field strength $B$. In particular, for  the proton cyclotron line we have
\be
\label{um9}
    E_{\rm 1} = 0.0635~ B_{\rm 13}~~ \rm keV.
\ee
For the description of the local
spectra of the model atmospheres above a condensed iron surface,
 two additional absorption features can be
considered. The first feature
is centered at the iron ion cyclotron energy according to
 Eq.\,(\ref{u2_0}). This
line is asymmetric, i.e., $\tau_{\rm 2} > 0$ only for $E > E_{\rm 2}$,
in agreement with the calculations
(see Figs.~\ref{f:fig3} and \ref{f:fig4}). The second line is centered
between $E_{\rm 1}$ and $E_{\rm C}$,
and its position depends also on the local magnetic field strength.

We use three angular distributions $\phi(\alpha)$ 
of the specific intensity $I_{\rm E}(\alpha)$.
The simplest one is the isotropic function $\phi(\alpha)$=1,
which corresponds to the thin model atmosphere above a
blackbody. For the semi-infinite model atmosphere the angular
distribution (\ref{u2_13}) are used. For the thin
atmosphere models above a condensed iron surface the isotropic
angular distribution is used, except for the energy band between
$E_{\rm C}$ and 4$E_{\rm C}$. Here the angular distribution has a
jump at some $\cos\alpha$ (see Fig.~\ref{f:fig6}), and this case
has to be studied separately.

For the description of the naked condensed surface spectra the relations (\ref{u2_4a}) -- (\ref{u2_7})
 are used.  

The total spectrum of the NS is calculated in the spherical coordinate system connected with the
rotation axis (see, for example, \citealt{PoutGerl:03} and references therein)
\be
\label{um12}
 f_{\rm E} = \frac{R^2}{d^{~2}} \int_0^{2\pi} d\,\varphi \int_0^{\pi} (1+z)^{-3}\, I_{\rm E'}(\alpha)
\,\cos\alpha\, \sin\gamma \,d\gamma,
\ee
where $d$ is the NS distance and
 $\gamma$ is the angle between the radius vector at a
 given point and the rotation axis. Here the observed
 and the emitted photon energies are connected by the relation $E=E'(1+z)^{-1}$,
and  $z = (1-R_{\rm S}/R)^{-1/2}-1$ is the gravitational redshift.
Light bending in the gravitational field
is accounted by the approximate relation \citep{Belob:02}
\be
\label{um13}
   \cos\alpha \approx \frac{R_{\rm S}}{R} + \frac{\cos\psi}{(1+z)^{2}},
\ee
where $R_{\rm S} = 2GM/c^2$ is the Schwarzschild radius, and
\be
\label{um14}
   \cos\psi = \cos i \cos\gamma + \sin i \sin\gamma\cos\varphi.
\ee
The integration in Eq.~(\ref{um12}) is restricted to
the visible NS surface area with $\cos\alpha > 0$.

Phase resolved spectra can be calculated for various phase angles, and the corresponding
light curves can also be calculated for various energy bands. The mean 
spectrum is summed from separate phase spectra and
then divided by the number of phases. 

\begin{figure}
\begin{center}
\includegraphics[angle=0,scale=0.6]{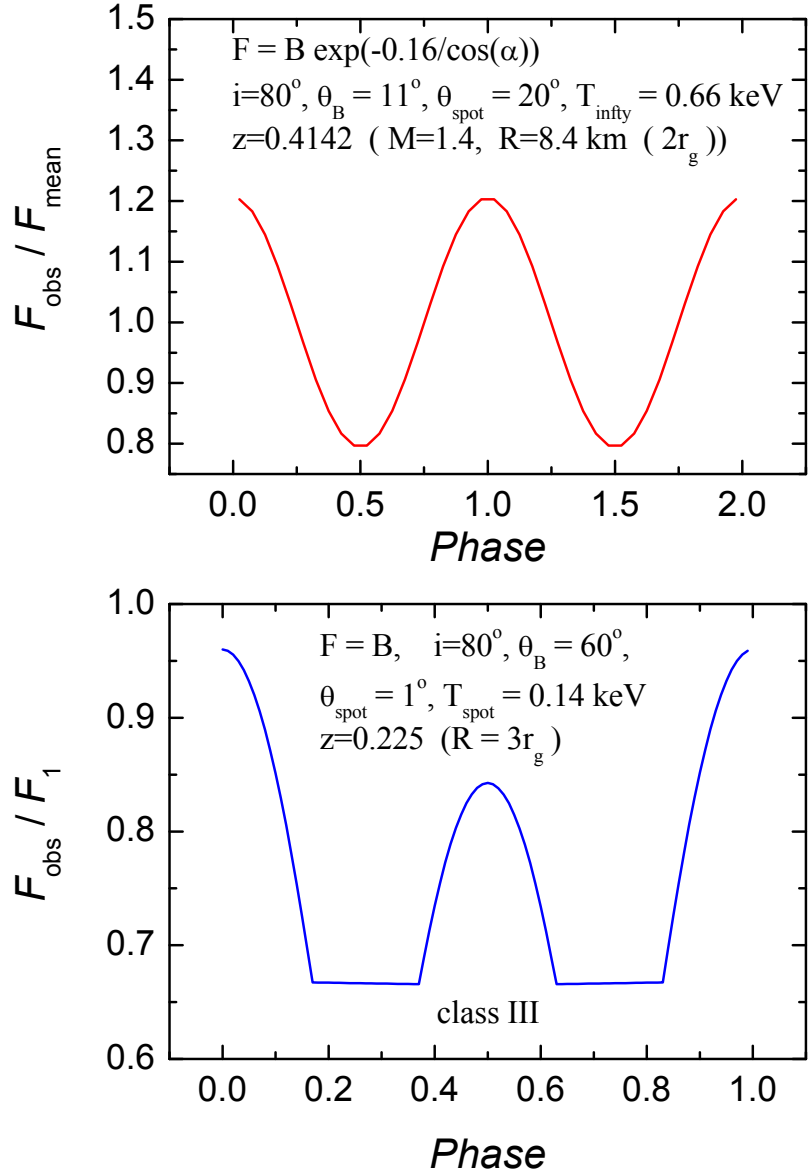}
\caption{\label{f:lctest}
 Test light curves calculated with the same parameters as used by
\citet{PoutGerl:03} (top panel) and \cite{Belob:02} (bottom panel).
$F_1$ is the maximum possible flux that is observed when one of the spots is
viewed face-on.
}
\end{center}
\end{figure}

 Using this code we calculated two test light curves from slowly rotating
  neutron stars (see Fig.~\ref{f:lctest}). The first one corresponds to the light
  curve presented by \citet{PoutGerl:03} in their Fig.\,4a (dashed curve), and the second one 
  corresponds to Class III light curves in the classification performed by
  \cite{Belob:02} (see his Fig.\,4).

\section {Results of modeling}
\label{s:results}

In this section we present some examples of model
light curves and integral emergent spectra for
magnetized neutron stars with various surface temperature distributions. We choose symmetric models
with equal temperatures and magnetic field strengths at both magnetic poles,
 and equivalent temperature and magnetic field
distributions in both hemispheres. We consider a neutron star model with polar temperature $kT_{\rm p} =
0.15$ keV, polar magnetic field strength $B_{\rm p} = 6 \times 10^{13}$ G and a compactness corresponding to
the gravitational redshift $z = 0.2$. Three temperature
distributions are used. Two of them are described by
Eq.\,(\ref{um4}) with $a=0.25$ and $a=60$.
In both distributions $T_{\rm min}$ equals 0.316$T_{\rm p}$.
 The results are almost the same for any smaller value of  $T_{\rm min}$.
The third temperature distribution is a uniform  surface temperature $kT = 0.01$ keV plus
two uniform bright spots around the magnetic poles with the same
temperatures $T_{\rm p}$ and angular radii
$\theta_{\rm sp} = 5\degr$. Most of the models are calculated for angles $i = \theta_{\rm B} = 90\degr$, which
gives the largest change of the visible local parameters during a rotation period. For simplicity we take dilution
$D$ and the hardness $f_c$ factors equal to 1.
 Most of our results will not be changed by this,
  in particular, it will not change the normalized spectra and EWs of absorption
lines, if $f_c$ and $D$ are constant over the neutron star surface. This approximation will
change the absolute value of the flux only. Therefore, it is important for 
 the ratio of X-ray to optical fluxes, because $f_c$ and $D$ can be
 different for X-ray and optical bands
(see Sect.\,4.3). 
Therefore, in this section we will consider the temperature distribution over
the neutron star as a color temperature distribution instead 
of the effective temperature distribution.

\begin{figure}
\begin{center}
\includegraphics[angle=0,scale=0.4]{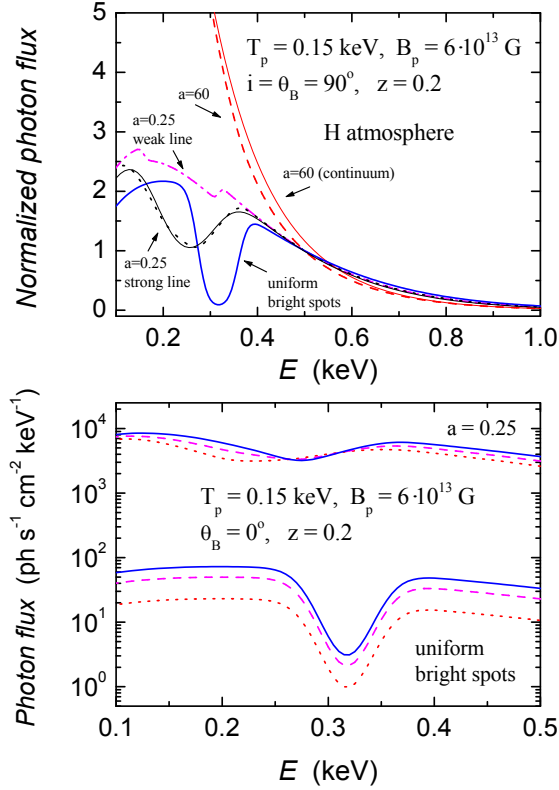}
\caption{\label{f:fig8}
Phase averaged photon spectra for
the neutron star model with $kT_{\rm
p1} = kT_{\rm p2} = 0.15$ keV, $B_{\rm p1} = B_{\rm p2} = 6
\times 10^{13}$ G, $z=0.2$, $i=\theta_{\rm B}= 90\degr$  and
different temperature distributions across the surface.
{\it Top panel:} the
temperature distribution Eq.\,(\ref{um4}) with $a_1=a_2=$ 0.25
(dotted curve),  with $a_1=a_2=$ 60 (dashed curve), and two
uniform bright spots with $T=T_{\rm p}$ and  angular size
$\theta_{\rm sp}$ = 5$\degr$ (solid curve). The local spectra
are isotropic blackbody spectra with an absorption line ($\tau_1
= 3$, $\sigma_1 = 30$ eV) centered according to Eq.\,(\ref{um9}).
 The spectrum of the model with $a=0.25$ and a
weaker  absorption line  ($\tau_1 = 2.6$, $\sigma_1 = 6$ eV) is
also shown (dash-dotted curve).  The thin solid curves are the
continuum spectrum of a model with $a=60$ and a single blackbody spectrum with a Gaussian line
($E_1 = 0.25$ keV, $\tau_1 = 0.85$, $\sigma_1 = 52$ eV) for the strong-line $a=0.25$ model.
{\it Bottom panel:} The photon spectra
for two temperature distributions
($a=0.25$ and two bright spots) under various inclination angles
to the line of sight: $i = 0\degr$ (solid curves), $i = 45\degr$
(dashed curves), and  $i = 90\degr$ (dotted curves). }
\end{center}
\end{figure}

\subsection{Averaged spectra}

The main goal of our modeling is the study of possible absorption features in the magnetized neutron star
thermal spectra. To this aim the spectra of the models with the three temperature distributions, as described above,
and the three local spectral energy distributions, as described in Section 2,
are computed.  In all the cases we consider the color temperature
  distribution over neutron star surface.

\subsubsection{Blackbody plus Gaussian line}

 The first spectral model is a simple isotropic
blackbody spectrum with a temperature equal to the local  temperature of a given model and
one Gaussian  spectral line. The line is centered at the proton cyclotron energy (see Eq.\,(\ref{um9})) and has
parameters $\tau_1 = 3$, $\sigma_1=30$ eV, hence EW $\approx 100$ eV. This model roughly  represents
the shape of the spectrum of a thin magnetized partially ionized hydrogen atmosphere above
the blackbody condensed surface (see Fig.~\ref{f:fig2}).
The same model can represent the shape of the spectrum of the semi-infinite
atmosphere at the maximum of the flux, but, of course, with larger
$f_c$. 
Model spectra of  semi-infinite magnetic atmospheres have hard tails in
   comparison with a blackbody. But to first approximation these spectra at
   the maximum flux can be  described as diluted blackbody spectra
   (see, for example, \citealt{sul:07} for non-magnetic atmospheres). Moreover,
   the model spectra with vacuum polarization and partial mode
   conversion have smaller deviations  from blackbody than the
    models, where these effects are ignored (see Fig.\,10 in
   \citealt{sul:09}). We mainly investigate here
the absorption features, and ignore the harder spectral tail in the real model
atmosphere spectra.

Integral photon spectra for all the
temperature distributions are shown in the top panel of Fig.~\ref{f:fig8}.
The overall shapes of the spectra can be described by blackbody spectra with
$kT \approx$ 0.114 keV for the model with $a=0.25$, $kT \approx$ 0.086 keV for the model with $a=60$,
and $kT \approx$ 0.125 keV for the model with two bright spots.
 Therefore, these spectra can be described in first approximation by single
  blackbody spectra with a Gaussian line.
The example of such an approximation is shown in Fig.~\ref{f:fig8}. The model spectrum with $a=0.25$ 
is well represented by the blackbody spectrum with $kT \approx$ 0.114 keV and a Gaussian line with $E_1 = 0.25$ keV, $\tau_1
= 0.85$, $\sigma_1 = 52$ eV. The integral spectra have a soft excess and an asymmetry of the line 
in comparison with single blackbody spectra.
These features increase if $a$ increases (see also Fig.\,\ref{f:fig11}).

 The spectrum of the model with  $a=60$ is
the softest one due to a relatively large contribution of the cooler parts of the neutron star.
 The spectrum of the two spots model has a relatively narrow absorption line,
because  most radiation comes from the pole regions where the magnetic field strength is nearly constant.
The spectrum of the model with $a=0.25$ shows a wide asymmetric line, shifted to lower energy.
 The large width of the line is a
consequence of the smooth temperature distribution, and,
accordingly, a large variety of magnetic field strengths, which
 effectively contribute to the spectrum.
 The absorption feature in the spectrum of the model with $a=60$ is very
 wide and difficult to see.
 It can be recognized by a comparison with the spectrum without
 the line (the thin curve in Fig.~\ref{f:fig8}, top panel).
In this model the magnetic field changes significantly, from $B_{\rm p}$ at the poles up to 7.7$B_{\rm p}$
at the equator and the absorption line is strongly smoothed. An observable absorption feature appears only in models
with  $a \le$ 10 -- 50 (depending on the EW of the
local absorption line).
Therefore, strong toroidal fields
on the surface of XDINSs (corresponding to $a >$  50) are incompatible
with the observed absorption lines.

 The equivalent widths of these absorption features range from 65
 eV ($a=60$ model) to 85 eV (the other models). Therefore,
  the magnetic field distribution does not result in an increase of the EW of the absorption line in
 the integral spectrum.
 For illustration the integral spectrum of the
model with $a=0.25$ and a line with $\tau_1 = 2.6$, $\sigma_1=6$ eV and EW $\approx 19$ eV
is also shown in the upper panel of Fig.~\ref{f:fig8}. The line  has the significant width but a
small depth with EW $\approx 17$~eV.
   
The integral spectra of the model with $a=0.25$ and the two-spots model at
different inclination angles  of the magnetic axis and rotation axis
  (which  coincide),
to the line of sight ($i=$ 0$\degr$, 45$\degr$ and 90$\degr$) are shown in the bottom panel of Fig.~\ref{f:fig8}.
These spectra represent phase resolved spectra at various phases of the stellar
rotation for the case $\theta_{\rm B} = 90{\degr}$. The spectra of the two-spots model only show the flux
level variation, without a change of the line shape. In contrast,
the spectra of the models with $a=0.25$ show line shape variations.
At larger $i$ the line center is
shifted to smaller energy and becomes wider, because a more significant part of the observed radiation comes from
the magnetic equator regions. The dependence of the absorption line EW on the inclination angle is insignificant.

\begin{figure}
\begin{center}
\includegraphics[angle=0,scale=0.6]{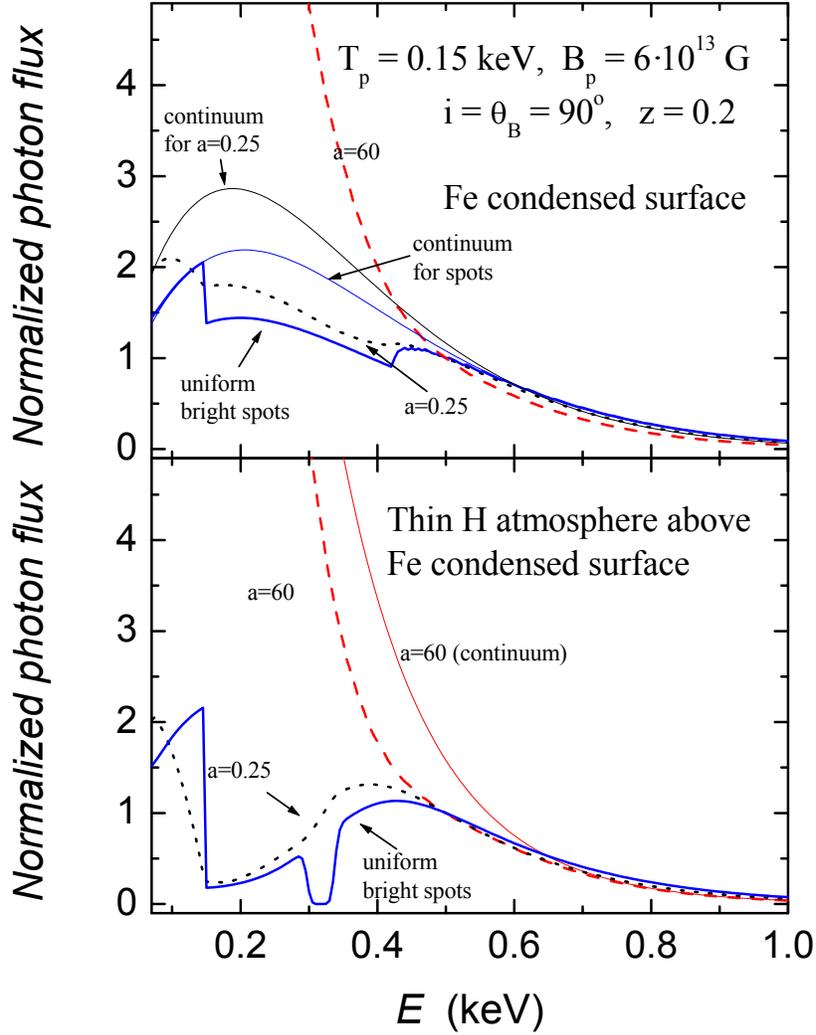}
\caption{\label{f:fig9}
{\it Top panel:} Phase averaged photon spectra of the same neutron star models as in Fig.~\ref{f:fig8}, but here the
 local spectra are from a naked iron surface. The continuum spectra without absorption 
features are shown by the thin solid curves
for the two-spots model and the model with $a=0.25$.
{\it Bottom panel:} Phase averaged photon spectra of the same neutron star models as in Fig.~\ref{f:fig8}, but here
the local spectra are isotropic blackbodies with absorption features like in the spectra of the thin hydrogen
model atmosphere above the condensed iron surface (see text). The continuum spectrum
 without absorption feature is shown by the thin solid curve
for the model with $a=60$.
 }
\end{center}
\end{figure}

\subsubsection{Naked condensed iron surface}

The second  model used for the local spectrum is the spectrum of
the naked condensed iron surface, provided by
Eqs.~(\ref{u2_4a}) -- (\ref{u2_7}). The  computed
spectra for the neutron star models with the same parameters and
temperature distributions, as in the case of the first local
model, are shown in the top panel of Fig.~\ref{f:fig9}.   These integral
spectra are close to blackbody spectra with similar color temperatures,
as in the first case (see Sect. 4.1.1).   The absorption features are very
wide and shallow and have  EWs equal to $\approx$ 110 eV for the
two-spots model, $\approx$ 140 eV for the model with $a=0.25$,
and  $\approx$ 95 eV for the model with $a=60$. In the last case
the line is again not observable, and  the EW is obtained with
a lower accuracy because of uncertainties in the continuum
definition. 

\begin{figure}
\begin{center}
\includegraphics[angle=0,scale=0.6]{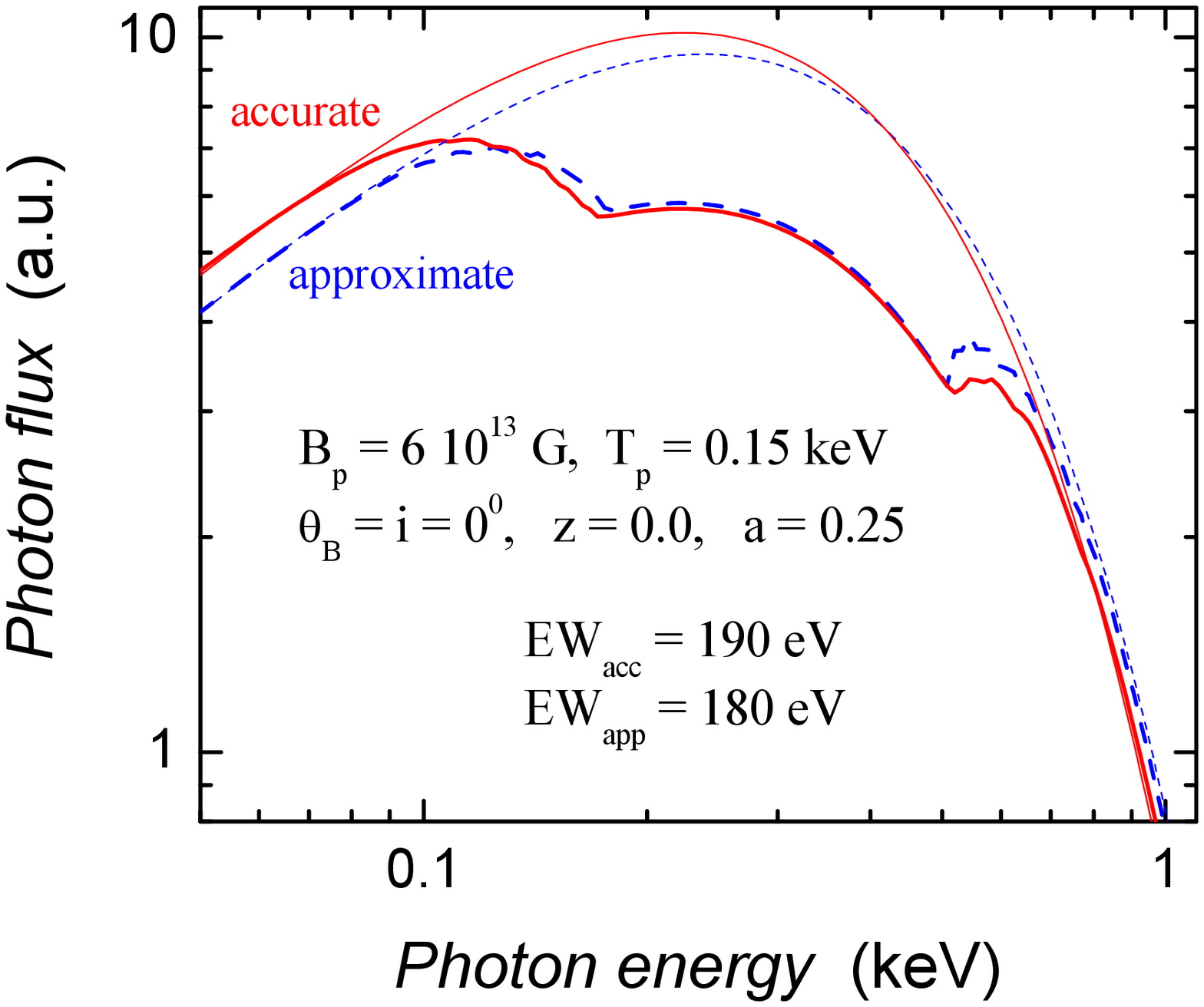}
\caption{\label{f:fig9a}
 Phase averaged photon spectra of the neutron star models with parameters $kT_{\rm p1}=kT_{\rm p2} = 0.15$ keV, 
$B_{\rm p1}=B_{\rm p2} = 6 \times 10^{13}$ G, $i = \theta_{\rm B} = 0^{\circ}$ and $a$ = 0.25.
The local spectra are from a naked iron surface, computed accurately using the code presented 
by \citet{vAetal:05} (thick solid surve), and using approximate Eqs. (\ref{u2_4a}) - (\ref{u2_7})
(thick dashed curve). The adopted continuum spectra without absorption 
features are shown by the corresponding thin curves. The EWs of the absorption features are rather similar,
$\approx$ 190 and 180 eV.
 }
\end{center}
\end{figure}

 It is possible to compare approximate calculations of neutron star spectra with naked condensed iron surface   
with the accurate calculations, performed by the method described in \citet{vAetal:05}.
An example of this comparison is shown in Fig.\,\ref{f:fig9a}. The qualitative agreement is satisfactory.
The accurate spectrum is (as expected) more smooth at $E_{\rm C}$, and the flux at energies less than
$E_{\rm c,i}$ is larger. The EWs of the absorption features are very similar, $\approx$ 190 eV for 
the accurate spectrum and  $\approx$ 180 eV for 
the approximate spectrum.

\subsubsection{Thin atmosphere above condensed iron surface}

 The shape of the spectrum of a thin hydrogen atmosphere above 
a condensed iron surface can be roughly fitted by a
blackbody spectrum with  two absorption features with Gaussian profiles, 
which describe the proton
cyclotron line and one half of the wide line at $E_{\rm c,i}$(Fe). 
The half
line means that the line optical depth is equal zero at energies 
$E < E_{\rm c,i}$(Fe), and calculated in the usual way
(Eq.\,(\ref{u2_9}))
at $E > E_{\rm c,i}$(Fe). For our illustrative calculations we
take $\tau_1 = 8.5$, $\sigma_1 = 11$ eV, and $\tau_2 = 2.5$,
$\sigma_2 = 150$ eV with a total EW $\approx$ 230 eV  (see also Fig.\,\ref{f:fig4}). The
Gaussian lines with these $\tau_i$ and $\sigma_i$ well fit the
absorption features in the  spectrum of the thin hydrogen 
atmosphere above the condensed iron surface with
$T_{\rm eff} = 1.2 \times 10^6$ K, $B = 10^{14}$ G, and $\Sigma =
10$ g cm$^{-2}$ (Fig.~\ref{f:fig4}). The computed spectra for
the used temperature distributions are shown in the bottom
panel of Fig.~\ref{f:fig9}. 
 Again, the temperature of the neutron star surface means the color temperature.
The hard tails of the integral spectra can be
fitted by the blackbody  spectra with approximately the same
temperatures as those obtained for the first local spectral model.
 The absorption feature has a prominent two-component
structure for the two-spots model,  and a smoothed shape for the
model with $a=0.25$. In the spectrum of the model with $a=60$ the
absorption line  is again very smooth. The EWs of the features
in spectra of the two-spot model and the model with $a=0.25$ are
$\approx$ 210 eV, and the EW of absorption feature in the
spectrum of the model with $a=60$ is $\approx$ 150 eV.

\begin{figure}
\begin{center}
\includegraphics[angle=0,scale=0.6]{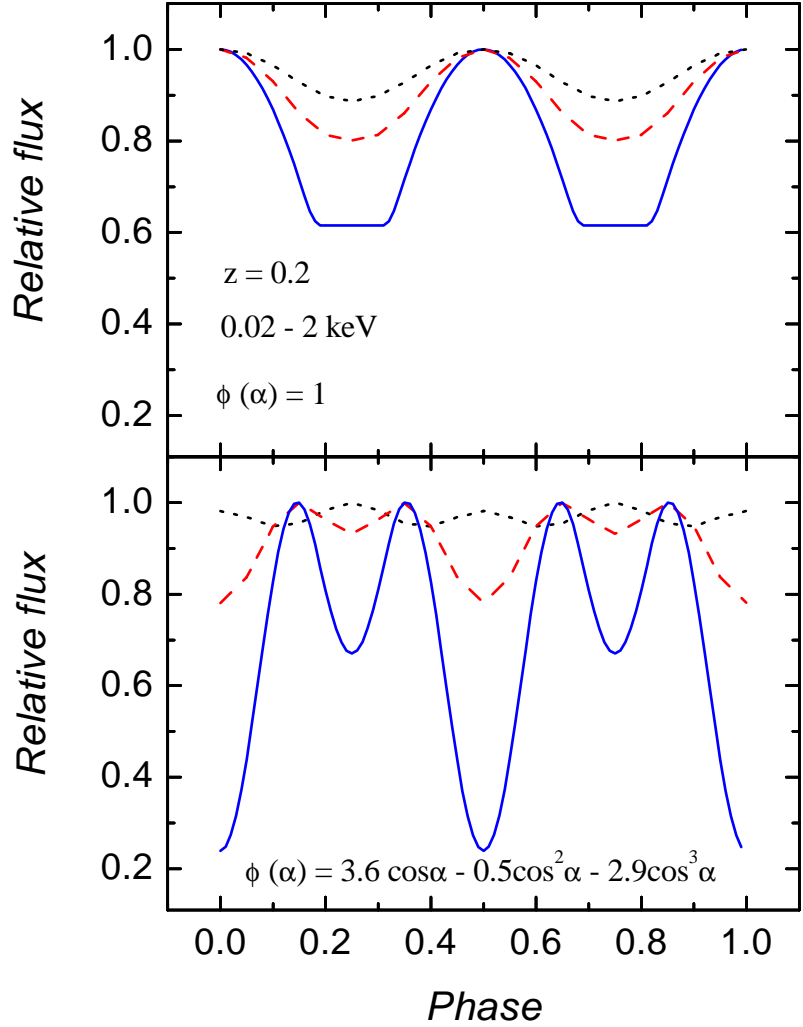}
\caption{\label{f:fig7}
Examples of light curves for the neutron star model with $kT_{\rm
p1} = kT_{\rm p2} = 0.15$ keV, $B_{\rm p1} = B_{\rm p2} = 6
\times 10^{13}$ G, $z=0.2$, $i=\theta_{\rm B}= 90\degr$  and
different temperature distributions across the surface: the
temperature distribution (\ref{um4}) with $a_1=a_2=$ 0.25
(dotted curves),  with $a_1=a_2=$ 60 (dashed curves), and two
uniform bright spots with $T=T_{\rm p}$ and  angular size
$\theta_{\rm sp}$ = 5$\degr$ (solid curves). Two different
angular distributions of the emergent radiation are used: the
isotropic distribution (top panel) and the semi-infinite magnetic
atmosphere distribution Eq.\,(\ref{u2_13}) (bottom panel).
 }
\end{center}
\end{figure}

\subsection{Light curves}

Light curves calculated for illustrative purposes in a
wide energy band (0.02--2 keV) are shown in
Fig.~\ref{f:fig7}. The light curves are computed for the
orthogonal rotator ($i = \theta_{\rm B} = 90\degr$), the three
temperature distributions, and two angular distributions of the
emergent flux: the isotropic one and the distribution corresponding
to a semi-infinite magnetized hydrogen model atmosphere as
described by  Eq.\,(\ref{u2_13}). The light curves, computed with
the isotropic angular distribution, agree with the well known fact
(see, e.g. \citealt{Page:95,Geppertal:06}) that the temperature
distribution corresponding to a core dipole magnetic field
($a=0.25$) cannot provide a pulsed fraction $PF\equiv (F_{\rm
max} - F_{\rm min})/ (F_{\rm max} + F_{\rm min})$
larger than several percent. In order to account for larger PF
values, more peaked temperature distributions have
to be considered \citep{Swopeetal:05}. Computed light curves
have $PF$ $\approx 5\%$ for the temperature distribution  with
$a=0.25$, $\approx 10\%$  for $a=60$, and $\approx 25\%$ for the
two bright spots.

 Using the angular distribution corresponding to the magnetized
model atmosphere, we obtain completely different pulse profiles
with four maxima instead of two for the isotropic angular
distributions  (bottom panel in Fig.~\ref{f:fig7}). The model
with $a=0.25$ shows additional maxima at the phases  of the
minima in the light curve computed with isotropic angular
distribution.  The models with the narrow peaked temperature
distributions (two spots and $a=60$) show deep minima at the
phases of the maxima in the light curves computed with isotropic
angular distribution. Each minimum in these light curves becomes
replaced by two maxima with a shallow minimum between them.  The
pulsed fractions of the light curves for the models with
$a=0.25$  and $a=60$ are the same as in the isotropic
distribution case, but the pulsed fraction of the light curve
for  the model with two bright spots increases up to 60\%.
 A qualitatively similar Class III light curve for a slowly rotating neutron star
  with  magnetic atmosphere model
  radiation  was obtained by \citet{Ho:07} (see his Fig.\,2).

\begin{figure}
\begin{center}
\includegraphics[angle=0,scale=0.4]{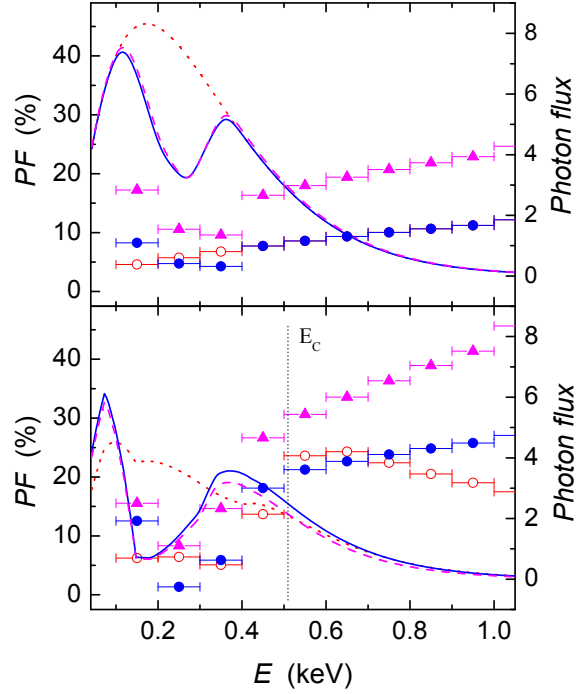}
\caption{\label{f:fig10}
Energy dependence of pulsed fractions (symbols and left vertical
axes) together with averaged photon spectra (lines and right
vertical axes) for the neutron star
models with the same parameters as in Fig.~\ref{f:fig8} and $a$=0.25.
{\it Top panel:}
The models with pure blackbody local spectra (open circles
and dotted curve), blackbody local spectra with an absorption line ($\tau$ = 3.0, $\sigma$=30 eV) centered
according to Eq.\,(\ref{um9}), and the
isotropic angular distribution (filled circles and solid curve), and blackbody local spectra with the
absorption line and the electron scattering atmosphere angular distribution (triangles and dashed curve).
{\it Bottom panel:} The model of a naked condensed iron surface
 (open circles
and dotted curve), a thin atmosphere above condensed iron surface spectra with corresponding angular distribution
(filled circles and solid curve), and the same local spectra, but with slightly different pole temperatures:
$T_{\rm p,1}$ = 0.15 keV, $T_{\rm p,2}$ = 0.14 keV (triangles and dashed curve). The position of $E_{\rm C}$, 
which corresponds to the magnetic pole, is shown by vertical dotted line.
 }
\end{center}
\end{figure}

We also investigate the influence of the absorption features on light curves and  pulsed fractions. This influence
on the light curves in a wide energy band is very small. But it is significant on the light curves and pulsed
fractions in the narrow energy bands at a vicinity of the absorption feature energies. For illustration we calculate the
dependences of $PF$ on the photon energy for the models with a smooth temperature distribution ($a$=0.25)
and the different local spectra (see Fig.~\ref{f:fig10}). The corresponding averaged spectra are presented
in the same figure for illustrative purpose. The common characteristic of the all presented dependences is the decreasing
of the pulsed fraction at the absorption feature energies.

In the top panel of Fig.~\ref{f:fig10} the local spectrum is  approximated by the blackbody
with and without an absorption line, and the changing in the pulsed fraction
at the absorption line energies very well seen.  The $PF$ for the bolometric light curve ($\sim 5\%$) is determined
by the soft part of the spectrum, where most of the photons come.
The $PF$ can be more significant for the higher energy
bands. We also calculate the averaged spectrum and the $PF$ dependence on energy using the same model but a more peaked
angular distribution of the local emergent radiation for an electron scattering atmosphere:
\be \label{ang}
  \phi(\alpha) = 0.4215+0.86775 \cos\alpha.
\ee
The averaged spectrum is changed insignificantly, but the pulsed fraction is increased by a factor of two
(the top panel in Fig.~\ref{f:fig10}).  We present this result in order to
  demonstrate that a pencil-like peaked angular distribution of
  the emergent radiation can give a similar effect like a peaked temperature distribution.

The dependence of the pulsed fraction on energy for the models with local spectra corresponding to a naked iron surface and a    
thin atmosphere above the condensed iron surface is similar (the bottom panel in Fig.~\ref{f:fig10}). The $PF$ at energies
$E > E_{\rm C}$ is larger due to peaked angular distributions (see Eqs.\,(\ref{u2_4a}) -- (\ref{u2_7}) and Fig.~\ref{f:fig6}).
 In our calculations we approximate the angular distribution of the thin atmosphere above the condensed surface 
emergent radiation by a simple step function in the $E_{\rm C}$ -- 4$E_{\rm C}$ energy range:
\be \label{add1}
\phi(\alpha) = \frac{1-a_{\mu}\cos\alpha^2}{1-\cos\alpha^2},~~~ {\rm if}~~~ \cos\alpha \ge \cos\alpha_{\rm c},
\ee
and
\be \label{add2}
\phi(\alpha) = a_{\mu},~~~ {\rm if}~~~ \cos\alpha < \cos\alpha_{\rm c},
\ee
where
\be \label{add3}
\cos\alpha_{\rm c}= \left(\frac{1}{3}\left(\frac{E_{\rm C_{\rm 1}}}{E_{\rm C}}-1\right)\right)^{2/3},
\ee
and $a_{\mu}$ is a parameter, which was taken 0.2 here. At other energies $\phi(\alpha)$=1 is used.

The pulsed fraction at these energies can be
even more significant if slightly different temperatures at the poles are used.
Here we do not take into account an azimuthal dependence of the
angular distribution of the magnetized  semi-infinite atmospheres
with inclined magnetic field. Therefore, the light curves can be
even more complicated. But the observed shapes of most XDINS
light curves are close to a simple sine, therefore we expect that
the angular  distribution of the emergent radiation from XDINS is
close to isotropic.

\subsection{Fluxes in the optical band}

\begin{figure}
\begin{center}
\includegraphics[angle=0,scale=0.4]{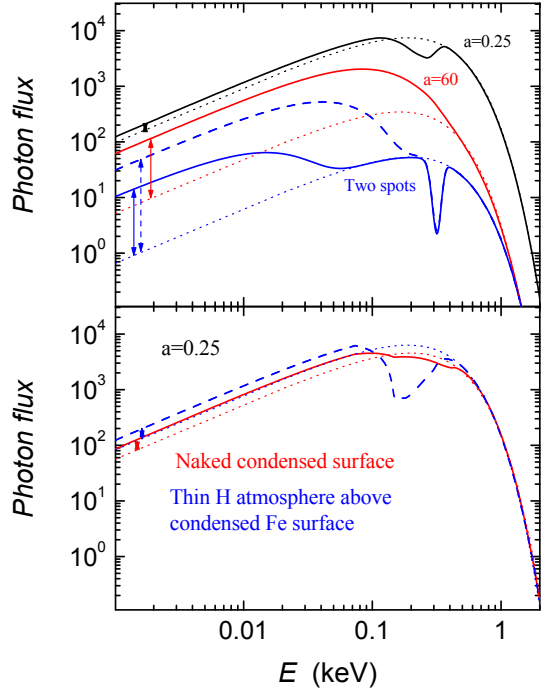}
\caption{\label{f:fig11}
Comparison of the integral averaged spectra of the magnetized neutron star models (solid and dashed curves)
with blackbody spectra, which fit the model spectra at $E \ge$ 0.5 keV (dotted curves). The differences at the optical
band are also shown.
{\it Top panel:} The models with blackbody local spectra and the Gauss line ($\tau$ = 3.0, $\sigma$=30 eV),
centered according to Eq.\,(\ref{um9}),
for three temperature distributions: $a$=0.25 (upper curves), $a$=60 (middle curves), and two-spot model 
(lower curves). The parameters of
the distributions are the same as in Fig.~\ref{f:fig7}. The spectrum of two-spot model with the 
temperature $kT$=0.03 keV of the cold surface 
is also shown (dashed curve).
{\it Bottom panel:} The models with smooth temperature distribution ($a$=0.25) and various local
 model spectra: naked iron spectrum
(solid ´curves) and the thin model atmosphere above condensed iron surface spectra (dashed curve). Parameters of the
local spectra are the same as in Fig.~\ref{f:fig9}.
 }
\end{center}
\end{figure}

Here we compare the fluxes of our computed models in the optical band with the fluxes of the blackbody extrapolations
of the X-ray model spectra. It is well known that the observed optical fluxes of XDINSs exceed the blackbody
extrapolations many times (see Introduction), therefore, this
comparison can give additional information on the
temperature distribution and the nature of local spectra. The comparisons are presented in Fig.\,\ref{f:fig11}. The models with
a smooth temperature distribution ($a$=0.25) show the small excesses  (factor $\approx$ 1.3 -- 1.5) for all local model spectra. 
The narrow peaked temperature distributions
show more significant excesses (up to a factor of $\sim10$), which depend on the values of $a$ and the adopted
 temperature for the cool part of the neutron star surface.

There is an additional uncertainty connected with dilution $D$
and hardness $f_{\rm c}$ factors. In our magnetized neutron  star
model spectra we represented the local model atmosphere spectra by
blackbody spectra with $D=f_{\rm c}=1$ for simplicity. But
in the real model atmosphere spectra these factors can differ
from 1 and, moreover, they can be different for the X-ray and the
optical band spectra \citep{pavlovetal:96,sul:09}.

\section{Conclusions}
\label{s:conclusions}

With this paper we have started our study of the XDINS surface
radiation properties. We have considered a few theoretical models
of a highly magnetized ($B > 10^{13}$ G) neutron star surface:
partially ionized hydrogen atmospheres, semi-infinite and thin;
condensed iron surfaces; and thin  partially ionized hydrogen
atmospheres above a condensed iron surface.

To describe the emergent spectra of the model atmospheres, we
have used the results by \cite{sul:09} and demonstrated that the
equivalent widths of the absorption  features do not exceed 100
eV. We have presented a simple analytical approximation of the
emergent spectra of the condensed surfaces, previously studied by
\cite{PerAzetal:05} and \cite{vAetal:05}, and evaluated the equivalent
widths of the absorption features in these spectra (120 -- 190 eV
for  XDINS magnetic fields, see Eq.\,(\ref{u2_3})).

 Using this approximation, we have studied thin partially
ionized hydrogen atmospheres above a condensed iron surface.
The radiation  properties of the condensed surface are used for the
inner boundary condition for the model atmospheres. Some examples
of such models are presented. We have demonstrated that
the interaction of the condensed surface radiation and the
radiation transfer in the atmosphere leads to strongly increased
equivalent widths of the absorption features, up to 300 -- 450
eV. Another interesting property found is the prominently peaked
angular distribution  of the radiation at energies between
$E_{\rm C}$ and 4$E_{\rm C}$. 

All of these different local models used to compute integral
emergent spectra of neutron stars with parameters close to XDINS ones.
We presented the modeling code for  light curves and integral spectra.
We have considered a magnetic field distribution with a possible toroidal component
and two different models for the temperature distribution on the neutron star surface. The first model is a simple
uniform relatively cold surface with two uniform hot spots  with given angular sizes at the magnetic poles.
The second model was taken from \cite{PerAzetal:06a} and describes a continuous temperature distribution,
peaked at the magnetic poles. An additional parameter $a$ (connected with a relative contribution of the
toroidal component of the magnetic field) controls the width of the temperature profile peaks.

Light curve modeling confirms that a classical core dipole
temperature distribution ($a=0.25$ in our model)  with local
blackbody spectra is not sufficient to explain the observed
pulsed fractions in XDINSs. Therefore, more peaked temperature
distributions   or more peaked angular distributions of the
specific intensity are needed. We have demonstrated that the
isotropic or peaked angular distribution
of the emergent
radiation, as well as angular distributions in the models of thin
atmospheres, better explains the observed pulse profiles in
comparison with the magnetized semi-infinite model atmosphere
angular distribution.

 We also investigated dependence of the pulsed fractions on energy. The pulsed fraction 
increases with photon energy due to a strong dependence of the thermal 
flux on the temperature in the Wien part of the spectrum.
The pulsed fraction decreases at absorption feature energies. The
 emergent flux angular distribution of a thin
atmosphere above the condensed iron surface and a naked condensed surface 
 is peaked to the surface normal at
an energy range between $E_{\rm C}$ and 4$E_{\rm C}$, therefore, the pulsed 
fraction is increased there. It is also
increased if the pole temperatures are different.

We computed some examples of integral emergent spectra of hot magnetized neutron star models
for various temperature distributions and all considered surface models. We show that the
absorption features in the integral spectra become wider, but do not show an increase of the equivalent widths due to
the variation of the magnetic field strength over the neutron star surface. Therefore, the observed EWs in XDINS
spectra have to be similar to the local EWs on the neutron star surfaces. Most of the XDINSs have observed
absorption lines with EWs $\le$ 100 eV, which can be explained by  magnetized model atmospheres
or naked condensed surfaces, but the
EW of the absorption line in the spectrum of RBS\,1223 is about 200 eV, and an atmosphere above the condensed
iron surface provides a better explanation.  

The models with a strong toroidal magnetic field on the neutron
star surface  ($a >$ 10 -- 50) display a too wide smoothed
absorption feature,  which cannot be detected by observations.
Therefore, a strong toroidal magnetic field component  on XDINSs
surfaces can be excluded.  At the same time the large pulsed
fraction in some XDINSs hint to the existence of small hot
regions on the neutron star surfaces. Therefore, the model with a
poloidal magnetic field concentrated  in the neutron star crust
\citep{Geppertal:04} and models with a strong toroidal component
in the crust,  which vanishes on the surface (for example,
\citealt{ag08}), can be applicable. On the other hand, 
similar result can be reached if the peaked
angular distribution of the emergent spectra is considered. For
example, the thin hydrogen atmosphere above the condensed iron
surface  with a smooth temperature distribution over the neutron star
surface ($a$=0.25) and slightly different pole temperatures can
provide the pulsed fraction observed  in RBS\,1223.

In the present work the radiation properties of the magnetized neutron star condensed surfaces are considered in a
rough approximation only. More accurate investigations of the magnetized model atmospheres above the
condensed surfaces are necessary in the future. However, our present model can be used for 
fitting of observed spectra of XDINSs to
obtain basic information about radiation properties of magnetized neutron star surfaces.

\begin{acknowledgements}

VS and VH acknowledge support by the \emph{Deut\-sche For\-schungs\-ge\-mein\-schaft (DFG)} through 
project C7 of SFB/Transregio 7 "Gravitational Wave Astronomy". VS also thanks the President's
programme for support of leading science schools (grant NSh-4224.2008.2) for partial financial support.
The work of AYP is supported by RFBR grant 08-02-00837  and the President's
programme for support of leading science schools (grant NSh-3769.2010.2).

\end{acknowledgements}

\end{document}